\documentclass[useAMS,usenatbib]{mn2e}
\usepackage{graphicx}
\usepackage{amssymb}
\def\lesssim{\!\!\!\phantom{\le}\smash{\buildrel{}\over
 {\lower2.5dd\hbox{$\buildrel{\lower2dd\hbox{$\displaystyle<$}}\over
                               \sim$}}}\,\,}
\def\mnras{MNRAS}
\def\apj{ApJ}
\def\apjl{ApJ}
\def\aj{AJ}
\def\araa{ARA\&A}
\def\aaps{A\&AS}
\def\aap{A\&A}
\def\pasp{PASP}
\def\apjs{ApJS}
\def\nat{Nat}
\def\procspie{Proc. SPIE}
\def\pasj{PASJ}
\def\baas{BAAS}

\def\CaII{Ca\,{\sc ii}} 
\def\HII{H\,{\sc ii}} 
\def\HeI{He\,{\sc i}}
\def\FeII{Fe\,{\sc ii}}
\def\KI{K\,{\sc i}} 
\def\OI{O\,{\sc i}}
\def\OII{O\,{\sc ii}}
\def\OIII{O\,{\sc iii}}
\def\MgI{Mg\,{\sc i}}
\def\NaI{Na\,{\sc i}}
\def\NaID{Na\,{\sc i}~D}
\def\ScII{Sc\,{\sc ii}}

\def\TiII{Ti\,{\sc ii}}

\title[SN 2009kn - the twin of SN 1994W]
{SN 2009kn - the twin of the Type IIn supernova 1994W}
\author[Kankare et al.]
{E. Kankare,$^{1,2}$\thanks{E-mail: erkki.kankare@utu.fi} M. Ergon,$^{3}$ F. Bufano,$^{4,5}$ J. Spyromilio,$^{6}$ S. Mattila,$^{1}$\and N. N. Chugai,$^{7}$ P. Lundqvist,$^{3}$ A. Pastorello,$^{5,8}$ R. Kotak,$^{8}$ S. Benetti,$^{5}$\and M.-T. Botticella,$^{5,8}$ R. J. Cumming,$^{9}$ C. Fransson,$^{3}$ M. Fraser,$^{8}$ G. Leloudas,$^{10,11}$\and M. Miluzio,$^{5}$ J. Sollerman,$^{3}$ M. Stritzinger,$^{12}$ M. Turatto$^{13}$ and S. Valenti$^{5,8}$\\
$^{1}$ Tuorla Observatory, Department of Physics and Astronomy, University of Turku, V\"ais\"al\"antie 20, FI-21500 Piikki\"o, Finland\\
$^{2}$ Nordic Optical Telescope, Apartado 474, E-38700 Santa Cruz de La Palma, Spain\\ 
$^{3}$ Oskar Klein Centre, Department of Astronomy, AlbaNova, Stockholm University, SE-10691 Stockholm, Sweden\\
$^{4}$ INAF - Osservatorio Astrofisico di Catania, via S. Sofia 78, I-95123 Catania, Italy\\
$^{5}$ INAF - Osservatorio Astronomico di Padova, Vicolo dell'Osservatorio 5, I-35122, Padova, Italy\\
$^{6}$ ESO, Karl-Schwarzschild-Strasse 2, 85748 Garching, Germany\\
$^{7}$ Institute of Astronomy of Russian Academy of Sciences, Pyatnitskaya St. 48, 119017 Moscow, Russia\\
$^{8}$ Astrophysics Research Center, School of Mathematics and Physics, Queen's University Belfast, BT7 1NN\\
$^{9}$ Onsala Space Observatory, Chalmers University of Technology, SE-43992 Onsala, Sweden\\
$^{10}$ Dark Cosmology Centre, Juliane Maries Vej 30, 2100 Copenhagen, Denmark\\ 
$^{11}$ Oskar Klein Centre, Department of Physics, Stockholm University, AlbaNova, SE-10691 Stockholm, Sweden\\
$^{12}$ Department of Physics and Astronomy, Aarhus University, Ny Munkegade 120, DK-8000 Aarhus C, Denmark\\
$^{13}$ INAF - Osservatorio Astronomico di Trieste, Via G. B. Tiepolo 11, I-34143, Trieste, Italy}

\date{}
\begin{document}

\maketitle

\label{firstpage}

\begin{abstract}
We present an optical and near-infrared photometric and spectroscopic study of supernova (SN) 2009kn spanning $\sim$1.5 yr from the discovery. The optical spectra are dominated by the narrow (full width at half-maximum $\sim1000$~km~s$^{-1}$) Balmer lines distinctive of a Type IIn SN with P Cygni profiles. Contrarily, the photometric evolution resembles more that of a Type IIP SN with a large drop in luminosity at the end of the plateau phase. These characteristics are similar to those of SN~1994W, whose nature has been explained with two different models with different approaches. The well-sampled data set on SN~2009kn offers the possibility to test these models, in the case of both SN~2009kn and SN~1994W. We associate the narrow P Cygni lines with a swept-up shell composed of circumstellar matter and SN ejecta. The broad emission line wings, seen during the plateau phase, arise from internal electron scattering in this shell. The slope of the light curve after the post-plateau drop is fairly consistent with that expected from the radioactive decay of $^{56}$Co, suggesting an SN origin for SN~2009kn. Assuming radioactivity to be the main source powering the light curve of SN~2009kn in the tail phase, we infer an upper limit for $^{56}$Ni mass of 0.023~M$_\odot$. This is significantly higher than that estimated for SN~1994W, which also showed a much steeper decline of the light curve after the post-plateau drop. We also observe late-time near-infrared emission which most likely arises from newly formed dust produced by SN~2009kn. As with SN~1994W, no broad lines are observed in the spectra of SN~2009kn, not even in the late-time tail phase. 

\end{abstract}


\section{Introduction}
It is widely accepted that stars more massive than $\sim8$~M$_{\odot}$ end their life cycles as core-collapse supernovae (CCSNe). To date, the more common CCSN Types Ib/c and plateau Type II (IIP) SNe have been studied extensively, but the narrow-line Type II (IIn) SNe remain less well studied. Such SNe are identified from strong and narrow hydrogen emission lines in the spectrum, especially the prominent H$\alpha$ line. For a review of SN classification, see \citet{filippenko97}. The narrow-line features of Type IIn SN spectra are thought to arise from the SN ejecta interacting with an H-rich circumstellar medium (CSM). Recently, \citet{li11} derived new relative rates for different SN subtypes, concluding that Type IIn SNe make up 5 per cent of all SNe in their volume-limited sample. 

Type IIn SNe exhibit the most heterogeneous properties among the different CCSN subtypes. However, \citet{kiewe12} proposed separating a subcategory of Type IIn SNe based on their slowly evolving narrow P Cygni lines [full width at half-maximum (FWHM) $\sim1000$~km~s$^{-1}$], exhibited especially by the H$\alpha$ line. \citet{chugai04} associated these events with SNe interacting with a massive CSM envelope ejected in an outburst rather than in a superwind. Following these authors, we identify a subclass of Type IIn SNe consisting of events similar to SN~1994W. Such objects include SN~1994ak \citep{filippenko97}, SN~1994W \citep[hereafter SCL98, CBC04, and  DHG09, respectively]{sollerman98, chugai04, dessart09}, SNe 1995G and 1999eb \citep{pastorello02}, SN~1999el \citep{dicarlo02}, SN~2005cl \citep{kiewe12}, SN~2011A \citep{pignata11} and SN~2011ht \citep{roming12}. In this work, we show that SN~2009kn also belongs to this subclass. The variety of other Type IIn SNe, such as SN~1988Z \citep{stathakis91, turatto93}, SN~1998S \citep{fassia01} and the ultraluminous SN~2006gy \citep[e.g.][]{smith10a}, makes them less easy to group into subclasses.

We present data on SN~2009kn obtained by a wide European SN collaboration\footnote{This is paper is based on European Southern Observatory (ESO) NTT long-term programme, in the framework of a large international collaboration for SN research. For the composition of the Collaboration and its scientific goals, we refer the reader to our web pages (http://graspa.oapd.inaf.it/).} comprising one of the most comprehensive data sets yet collected for this subclass of Type IIn SNe. In Section~2, we give the basic parameters of SN~2009kn. In Sections~3 and 4, the photometric and spectroscopic data are described. In Section~5, we discuss the intrinsic nature of SN~2009kn and compare it to SN~1994W. A summary is given in Section~6. 

\section{SN~2009\MakeLowercase{kn}}
\citet{gagliano09} discovered SN~2009kn on 2009 October 26.46 {\sc ut} (JD~$=2455130.96$) in the Sb-type spiral galaxy MCG$-$03-21-006 (also known as ESO 561-G020) by comparison to their reference images obtained in 2005. Spectroscopic observations by \citet{steele09} on November 9 revealed it to be a Type IIn SN based on the detection of narrow spectral Balmer lines. A radio non-detection on November 27.43 {\sc ut} was reported by \citet{chandra09}. 

The recession velocity of 4931~km~s$^{-1}$ of MCG$-$03-21-006 \citep{chamaraux99}, corrected for Virgo and Great Attractor peculiar motion \citep{mould00}, gives a heliocentric redshift $0.015798 \pm 0.000033$. The corresponding luminosity distance is 70.3~Mpc (H$_{0}=71$~km~s$^{-1}$Mpc$^{-1}$, $\Omega_{\Lambda}=0.7$ and $\Omega_{\rm M}=0.3$) and the distance modulus $\mu=34.23$~mag. 

The lack of pre-discovery imaging of the field of SN~2009kn close in time to the SN discovery prevents us from establishing an accurate date of explosion. However, by cross-correlating the spectra of SN~2009kn to the spectral sequence of SN~1994W, we obtain an estimate of the explosion date of JD~$=2455116_{-11}^{+10}$ (October 11). We adopt this as the explosion date in the remainder of the paper. See Section~4.2 for details and caveats on the method. 

The spectra of SN~2009kn show no \NaI\ absorption at the host galaxy redshift. We therefore assume that the host galaxy extinction is negligible compared to the Galactic component. We adopt the Galactic extinction of $A_V=0.353$~mag as the extinction towards SN~2009kn (see Section~4.2 for details). 

\section{Photometry}

Photometric observations of SN~2009kn were obtained with the 2.56-m Nordic Optical Telescope \citep[NOT;][]{djupvik10} with the Andalucia Faint Object Spectrograph and Camera (ALFOSC\footnote{The data presented here were obtained in part with ALFOSC, which is provided by the Instituto de Astrofisica de Andalucia (IAA) under a joint agreement with the University of Copenhagen and NOTSA.}) and the Nordic Optical Telescope near-infrared Camera and spectrograph (NOTCam), the 3.56-m New Technology Telescope (NTT) with the ESO Faint Object Spectrograph and Camera v.2 \citep[EFOSC2;][]{buzzoni84} and the Son of ISAAC \citep[SofI;][]{moorwood98}, the 2.2-m Calar Alto (CA) telescope with the Calar Alto Faint Object Spectrograph \citep[CAFOS;][]{meisenheimer98}, the 2.0 meter Liverpool Telescope \citep[LT;][]{steele04} with the RATCam and the SupIRCam, and the 2 $\times$ 8.4 m Large Binocular Telescope (LBT) with the LBT Near Infrared Spectroscopic Utility with Camera and Integral Field Unit for Extragalactic Research \citep[LUCIFER;][]{mandel00}. We report on 22 epochs of optical imaging from day 15 to day 446 post-explosion and 12 epochs of NIR imaging from day 76 to day 531. 

\subsection{Data reduction}

The basic data reduction, i.e. bias subtraction, flat-fielding, alignment and co-addition of the exposures, were carried out using standard Image Reduction and Analysis Facility ({\sc iraf})\footnote{{\sc iraf} is distributed by the National Optical Astronomy Observatories, which are operated by the Association of Universities for Research in Astronomy, Inc., under cooperative agreement with the National Science Foundation.} tasks. The sky subtraction for the near-infrared (NIR) data was done using the {\sc xdimsum}\footnote{Experimental Deep Infrared Mosaicing Software} package developed for {\sc iraf}. 

\begin{figure}
\begin{minipage}{85mm}
\includegraphics[width=85mm, clip]
{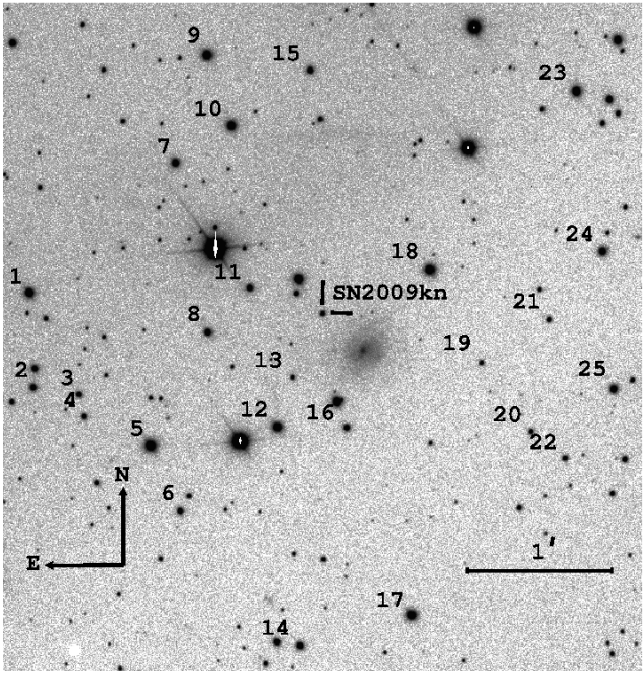}
\caption{\textit{R}-band $6.3 \times 6.3$~arcmin$^{2}$ NOT/ALFOSC image of the SN~2009kn field, JD~$=2455171.6$. North is up and east to the left.}
\label{fc}
\end{minipage}
\end{figure}

Instrument-specific optical colour terms were derived using several epochs of standard field observations. The calibration of the optical photometry was done relative to a sequence of up to 25 local stars, in turn calibrated using the photometric standard star field RU 152 \citep{landolt92} observed on JD~$=2455171.6$ with the NOT. The night was later determined as photometric based on the comparison of the derived \textit{UBVRI} zero-points to the instrumental zero-points reported\footnote{http://www.not.iac.es/instruments/alfosc/zpmon/} for NOT/ALFOSC and the atmospheric extinction monitoring measurements at the Carlsberg Meridian Telescope.\footnote{http://www.ast.cam.ac.uk/$\sim$dwe/SRF/camc\textunderscore extinction.html} SN~2009kn and its comparison stars are shown in Fig.~\ref{fc}. The calibration of the NIR photometry used up to 20 field stars from the Two Micron All Sky Survey (2MASS)\footnote{http://www.ipac.caltech.edu/2mass/index.html} \citep{skrutskie06}. The optical magnitudes for the field stars are reported in Table~1, and for completeness the 2MASS magnitudes are included. We carried out photometry of SN~2009kn using a point spread function (PSF) procedure based on the {\sc daophot} package in {\sc iraf}. A polynomial surface fit to the square background region surrounding the object position was derived and subtracted. The local stars used for photometric calibration were used to derive a PSF for each image which was then fitted to the SN. Measurement errors were estimated by simulating and fitting nine artificial PSF sources around the SN-subtracted residual image. This error usually dominated over the statistical uncertainty of the PSF fitting. The total error we report is the quadratic sum of the aforementioned measurement error and the standard error of mean of the zero-point values derived from individual local stars used for the calibration. Most of the {\sc iraf} tasks were run under the {\sc quba} pipeline.\footnote{Python package specifically designed by SV for SN imaging and spectra reduction. For more details on the pipeline, see \citet{valenti11}.} The optical photometry of SN~2009kn is reported in Table~2, and the NIR photometry in Table~3.

\begin{table*}
\begin{center}
\caption{Magnitudes of the SN~2009kn field stars (for the identifications, see Fig.~\ref{fc}). The 1$\sigma$ statistical errors are given in brackets.}
\begin{tabular}{ccccccccc}
\hline
 Star & $m_{U}$ & $m_{B}$ & $m_{V}$ & $m_{R}$ & $m_{I}$ &$m_{J}$ & $m_{H}$ & $m_{K}$ \\
 \# & (mag) & (mag) & (mag) & (mag) & (mag) & (mag) & (mag) & (mag)  \\ \hline
1 & 14.722(0.005) & 14.852(0.011) & 14.313(0.026) & 13.958(0.025) & 13.717(0.049) & 13.143(0.026) & 12.920(0.031) & 12.885(0.036) \\
2 & 16.063(0.012) & 16.213(0.014) & 15.652(0.027) & 15.186(0.026) & 14.919(0.044) & 14.491(0.034) & 14.243(0.047) & 14.156(0.082) \\
3 & 17.224(0.021) & 17.198(0.016) & 16.510(0.022) & 16.015(0.024) & 15.694(0.040) & 15.249(0.057) & 14.958(0.083) & 14.747(0.129) \\
4 & 17.388(0.024) & 17.196(0.015) & 16.455(0.023) & 15.899(0.026) & 15.562(0.041) & 15.015(0.044) & 14.525(0.056) & 14.530(0.096) \\
5 & 14.388(0.006) & 14.225(0.007) & 13.546(0.045) & 13.318(0.058) & 12.977(0.196) & 12.321(0.023) & 12.033(0.026) & 12.004(0.023) \\
6 & 16.615(0.016) & 16.655(0.012) & 15.960(0.017) & 15.492(0.026) & 15.131(0.042) & 14.664(0.037) & 14.369(0.049) & 14.324(0.072) \\
7 & 15.489(0.008) & 15.525(0.006) & 14.867(0.008) & 14.544(0.012) & 14.256(0.027) & 13.735(0.028) & 13.514(0.030) & 13.516(0.043) \\
8 & 16.586(0.014) & 16.168(0.008) & 15.380(0.013) & 14.947(0.008) & 14.592(0.026) & 14.077(0.028) & 13.782(0.033) & 13.767(0.057) \\
9 & 15.010(0.007) & 14.876(0.011) & 14.185(0.011) & 13.742(0.036) & 13.433(0.050) & 12.943(0.023) & 12.654(0.025) & 12.566(0.033) \\
10 & 14.669(0.006) & 14.762(0.008) & 14.223(0.010) & 13.888(0.023) & 13.647(0.048) & 13.177(0.025) & 12.938(0.026) & 12.907(0.030) \\
11 & 16.181(0.012) & 16.136(0.007) & 15.525(0.009) & 15.204(0.007) & 14.926(0.022) & 14.460(0.028) & 14.220(0.039) & 14.007(0.060) \\
12 & 15.107(0.006) & 15.046(0.005) & 14.345(0.018) & 13.970(0.013) & 13.644(0.050) & 13.053(0.023) & 12.668(0.027) & 12.660(0.030) \\
13 & 17.828(0.038) & 17.981(0.020) & 17.442(0.017) & 17.110(0.012) & 16.744(0.030) & - & - & - \\
14 & 15.806(0.012) & 15.959(0.015) & 15.506(0.009) & 15.188(0.035) & 14.848(0.051) & 14.631(0.035) & 14.387(0.045) & 14.279(0.075) \\
15 & 16.208(0.013) & 16.237(0.011) & 15.685(0.010) & 15.385(0.025) & 15.110(0.042) & 14.574(0.037) & 14.273(0.043) & 14.419(0.103) \\
16 & 16.554(0.016) & 16.522(0.008) & 15.891(0.015) & 15.533(0.009) & 15.125(0.030) & 14.711(0.038) & 14.373(0.054) & 14.166(0.075) \\
17 & 14.933(0.009) & 14.866(0.012) & 14.175(0.010) & 13.832(0.052) & 13.525(0.069) & 13.019(0.025) & 12.704(0.026) & 12.594(0.032) \\
18 & 14.814(0.007) & 14.683(0.006) & 14.080(0.008) & 13.701(0.029) & 13.385(0.036) & 12.905(0.026) & 12.643(0.025) & 12.541(0.032) \\
19 & 18.566(0.069) & 18.046(0.021) & 17.196(0.013) & 16.756(0.018) & 16.286(0.016) & - & - & - \\ 
20 & 17.727(0.037) & 17.625(0.015) & 16.891(0.012) & 16.539(0.014) & 16.050(0.019) & - & - & - \\ 
21 & 17.487(0.028) & 17.463(0.015) & 16.850(0.014) & 16.534(0.020) & 16.155(0.024) & - & - & - \\ 
22 & 17.822(0.037) & 17.628(0.016) & 16.898(0.011) & 16.584(0.017) & 16.104(0.018) & - & - & - \\ 
23 & 16.545(0.018) & 15.644(0.010) & 14.660(0.026) & 14.070(0.056) & 13.442(0.086) & 12.685(0.023) & 12.157(0.027) & 12.078(0.028) \\
24 & 16.149(0.015) & 15.747(0.006) & 15.013(0.012) & 14.593(0.030) & 14.202(0.044) & 13.477(0.025) & 13.029(0.029) & 12.977(0.032) \\
25 & 16.069(0.015) & 15.739(0.007) & 15.024(0.007) & 14.702(0.023) & 14.294(0.035) & 13.727(0.026) & 13.374(0.033) & 13.298(0.041) \\
\hline
\end{tabular}
\end{center}
\end{table*} 

\begin{table*}
\begin{center}
\caption{Optical photometry of SN~2009kn. The errors are given in brackets.}
\begin{tabular}{cccccccc}
\hline
JD & Epoch & $m_{U}$ & $m_{B}$ & $m_{V}$ & $m_{R}$ & $m_{I}$ & Telescope\\
(2400000+) & (d) & (mag) & (mag) & (mag) & (mag) & (mag) & \\ \hline

55131.0 & 15 & - & - & - & 16.623(0.164) & - & Puckett$^{a}$\\
55140.0 & 24 & - & - & - & 16.407(0.533) & - & Puckett\\
55153.0 & 37 & - & - & - & 16.806(0.491) & - & Puckett\\
55161.7 & 46 & 16.612(0.018) & 17.281(0.027) & 16.995(0.009) & 16.705(0.014) & 16.494(0.015) & NOT\\
55171.6 & 56 & 17.021(0.020) & 17.515(0.009) & 17.139(0.011) & 16.843(0.014) & 16.633(0.015) & NOT\\
55176.6 & 61 & 17.255(0.026) & 17.616(0.013) & 17.211(0.013) & 16.911(0.015) & 16.666(0.017) & NOT\\
55189.8 & 74 & 18.028(0.024) & 18.061(0.022) & 17.526(0.025) & 17.168(0.031) & 16.882(0.024) & NTT\\
55193.6 & 78 & - & 18.159(0.071) & 17.566(0.053) & 17.177(0.062) & 17.004(0.077) & CA\\
55204.6 & 89 & 18.906(0.069) & 18.551(0.059) & 17.935(0.019) & 17.550(0.022) & 17.220(0.019) & NOT\\
55209.6 & 94 & - & 18.882(0.016) & 18.052(0.017) & 17.646(0.013) & 17.326(0.013) & LT\\
55215.5 & 100 & 20.350(0.483) & 19.186(0.129) & 18.435(0.254) & 17.679(0.172) & 17.495(0.031) & NOT\\
55216.6 & 101 & - & 19.314(0.055) & 18.361(0.014) & 17.882(0.015) & 17.531(0.013) & LT\\
55219.7 & 104 & 20.860(0.100) & 19.632(0.025) & 18.549(0.022) & 18.044(0.043) & 17.671(0.045) & NTT\\
55224.5 & 109 & - & 20.384(0.182) & 19.335(0.143) & 18.989(0.055) & 18.620(0.046) & LT\\
55231.5 & 116 & - & 21.996(0.099) & 20.791(0.135) & 20.083(0.063) & 19.814(0.042) & NOT\\
55237.5 & 122 & - & 21.981(0.043) & 20.940(0.033) & 20.208(0.042) & 19.895(0.029) & LT\\
55245.7 & 130 & - & 22.135(0.081) & 21.009(0.036) & 20.307(0.036) & 19.887(0.028) & NTT\\
55260.7 & 145 & - & 22.212(0.062) & 21.139(0.038) & 20.445(0.056) & 19.928(0.064) & NTT\\
55273.4 & 157 & - & 22.483(0.040) & 21.244(0.040) & 20.686(0.030) & 20.153(0.027) & NOT\\
55296.4 & 180 & - & 22.739(0.085) & 21.541(0.096) & 20.904(0.069) & 20.200(0.140) & NOT\\
55475.9 & 360 & - & - & - & 21.974(0.102) & 21.576(0.183) & NTT\\
55561.6 & 446 & - & - & $>$22.675 & $>$21.918 & 21.968(0.491) & NTT\\
\hline
\end{tabular}
\end{center}
\begin{flushleft}
$^{a}$ We calibrated the Puckett Observatory unfiltered data in a similar way to the \textit{R}-band photometry of SN~2009kn.
\end{flushleft}
\end{table*}

\begin{table*}
\begin{center}
\caption{NIR photometry of SN~2009kn. The errors are given in brackets.}
\begin{tabular}{cccccc}
\hline
JD & Epoch &$m_{J}$ & $m_{H}$ & $m_{K}$ & Telescope\\
(2400000+) & (d) & (mag) & (mag) & (mag) & \\ \hline
55191.7 & 76 & 16.425(0.026) & 16.172(0.037) & 15.961(0.049) & NTT\\
55209.6 & 94 & 16.883(0.041) & 16.661(0.041) & - & LT\\
55212.8 & 97 & - & - & 16.417(0.047) & LBT\\
55219.5 & 104 & 17.107(0.039) & 16.984(0.044) & 16.481(0.049) & NOT\\
55244.8 & 129 & - & 18.842(0.032) & 18.470(0.034) & LBT\\
55246.7 & 131 & 19.208(0.061) & 18.871(0.131) & 18.390(0.150) & NTT\\
55261.7 & 146 & 19.430(0.072) & 19.095(0.114) & 18.660(0.139) & NTT\\
55305.6 & 190 & 19.665(0.202) & 19.248(0.129) & - & NTT\\
55315.4 & 199 & 19.838(0.159) & - & - & NOT\\
55532.9 & 417 & 21.162(0.121) & 19.633(0.102) & 18.414(0.046) & LBT\\
55605.6 & 490 & - & 19.987(0.165) & 18.668(0.109) & NTT\\
55646.6 & 531 & - & 20.055(0.141) & 18.904(0.120) & NTT\\
\hline
\end{tabular}
\end{center}
\end{table*} 

The local stars were also used to derive the World Coordinate System (WCS) solution for the field. This yielded RA~$=08^{\mathrm{h}}09^{\mathrm{m}}43.044^{\mathrm{s}}$ and Dec~$=-17^{\circ}44\arcmin51.12\arcsec$ (J2000.0) for SN~2009kn, $17.75$~arcsec east and $15.27$~arcsec north of the \textit{R}-band nucleus of the host galaxy MCG$-$03-21-006, corresponding to a projected distance of 8~kpc. Our coordinates differ slightly from those measured by \citet{gagliano09}. We believe that the larger discrepancy in RA is due to a typo or an error in converting angles to seconds of RA. The host galaxy has an inclination of $42^{\circ}$ according to HyperLeda \citep{paturel03}.

From the late epochs of our photometry, it is evident that the SN is projected close to a point-like source, most likely an \HII~region in the host galaxy. For the photometry of these epochs, the PSF was fitted both to the SN and to the nearby source to prevent it from affecting the PSF fitting. We obtain similar magnitudes of this nearby source at different epochs: $m_{B}=22.37\pm0.18$~mag, $m_{V}=22.18\pm0.09$~mag, $m_{R}=21.98\pm0.12$~mag, and $m_{I}=22.11\pm0.10$~mag. This suggests that the source does not strongly contaminate our late time photometry. During the early plateau phase, the SN is much brighter than this source, which therefore can be ignored.

\subsection{Light curves}

\begin{figure}
\begin{minipage}{85mm}
\includegraphics[width=85mm, clip]
{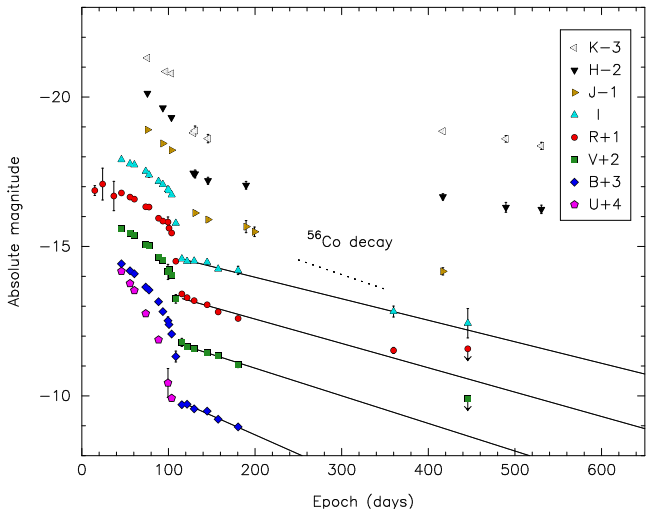}
\caption{Multiband absolute magnitude light curves of SN~2009kn. The slope expected from radioactive decay is indicated by the dotted line. The first three \textit{R}-band points of photometry were derived directly from unfiltered amateur observations of \citet{gagliano09} and are shown for completeness. The symbols are for most points larger than the error bars. Possible errors from determining the distance and extinction have not been propagated. The solid lines show fitted slopes for the tail phase decline rates in \textit{BVRI} bands. See text for details.}
\label{lc}
\end{minipage}
\end{figure}

All our multiband photometric follow-up observations have been conducted after maximum light, starting from 46 d after our adopted explosion date JD~$=2455116$. The absolute \textit{UBVRIJHK} light curves are shown in Fig.~\ref{lc}. The light curves show a declining plateau phase extending $\sim$100 d from the explosion date. The plateau resembles those shown by Type IIP SNe, which exhibit a distinctive hydrogen recombination phase in their light curves. Assuming a continuous plateau before the day 46 multiband observations, we estimate that the SN absolute peak magnitude was roughly $M_{B}\approx-18$~mag. Compared to the average peak magnitudes of Type II SNe \citep{richardson02}, SN~2009kn is $\sim$1~mag brighter than the Type IIP SNe and $\sim$1~mag fainter than the Type IIn SNe in their sample. Nevertheless, both Type IIP and Type IIn peak magnitude distributions show a wide spread and SN~2009kn cannot be excluded from either of the two classes based on the early photometry. SN~2009kn's plateau was followed by a rapid drop of $\sim$2.4~mag in \textit{B} and $\sim$2.0~mag in \textit{V}, \textit{R}, and \textit{I} in just 12 d, similar to SN~1994W which faded by $\sim$3.5~mag in \textit{V} in 12 d (SCL98). From day 116 onwards the decline rates, measured using least-squares fitting, were $\gamma_{B}=1.29\pm0.12$~mag (100 d)$^{-1}$, $\gamma_{V}=0.92\pm0.11$~mag (100 d)$^{-1}$, $\gamma_{R}=0.81\pm0.04$~mag (100 d)$^{-1}$, and $\gamma_{I}=0.72\pm0.06$~mag (100 d)$^{-1}$. Here we have included the measured magnitudes in \textit{R} and \textit{I} on day 360 and 446 in the fit, but not the upper limits in \textit{V} and \textit{R}. The slope in \textit{V} is comparable to the slope of 0.98~mag (100 d)$^{-1}$ expected from the radioactive decay of $^{56}$Co to $^{56}$Fe, assuming complete $\gamma$-ray and e$^{+}$ trapping. The slopes in \textit{R} and \textit{I}, however, are not as steep. This could indicate ongoing CSM interaction which enhances the flux in the H$\alpha$ and \CaII\ triplet lines in \textit{R} and \textit{I} bands, respectively. For more details, see Sections~4 and 5. See Section~3.3 for an analysis of the decline of the bolometric light curve. The gradual slope after the fast drop in the light curve is one of the main differences between SN~2009kn and SN~1994W, the latter exhibiting a much steeper decline rate. 

\begin{figure}
\begin{minipage}{85mm}
\includegraphics[width=85mm, clip]
{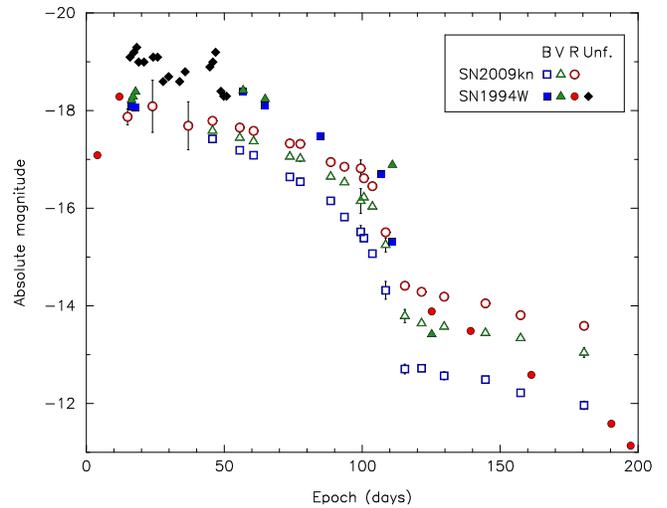}
\caption{Comparison of the absolute \textit{BVR} light curves of SN~2009kn and SN~1994W. The first three epochs of photometry of SN~2009kn are derived directly from unfiltered images and shown for completeness as in Fig.~1. The SN~1994W photometry (SCL98) is complemented with the unfiltered photographic photometry obtained by amateur astronomers (SCL98, and references therein) and corrected for \textit{B}-band extinction.}
\label{bvr}
\end{minipage}
\end{figure}

In Fig.~\ref{bvr}, the absolute \textit{VRI} light curves of SN~2009kn and SN~1994W are compared. The available photometry of SN~1994W between days 20 and 50 consists only of amateur photometry, which were converted to \textit{V}-band magnitudes by SCL98. SN~1994W exhibited a peak in the light curve at around 20$-$30~d from the date of explosion, after which the light curve remained on a plateau-like shape. For SN~2009kn, we are similarly missing early epoch photometry. We have derived \textit{R}-band magnitudes from the available unfiltered amateur images of SN~2009kn, shown in Fig.~\ref{lc}, but no clear light-curve peak is seen.

\begin{figure*}
\includegraphics[width=170mm, clip]
{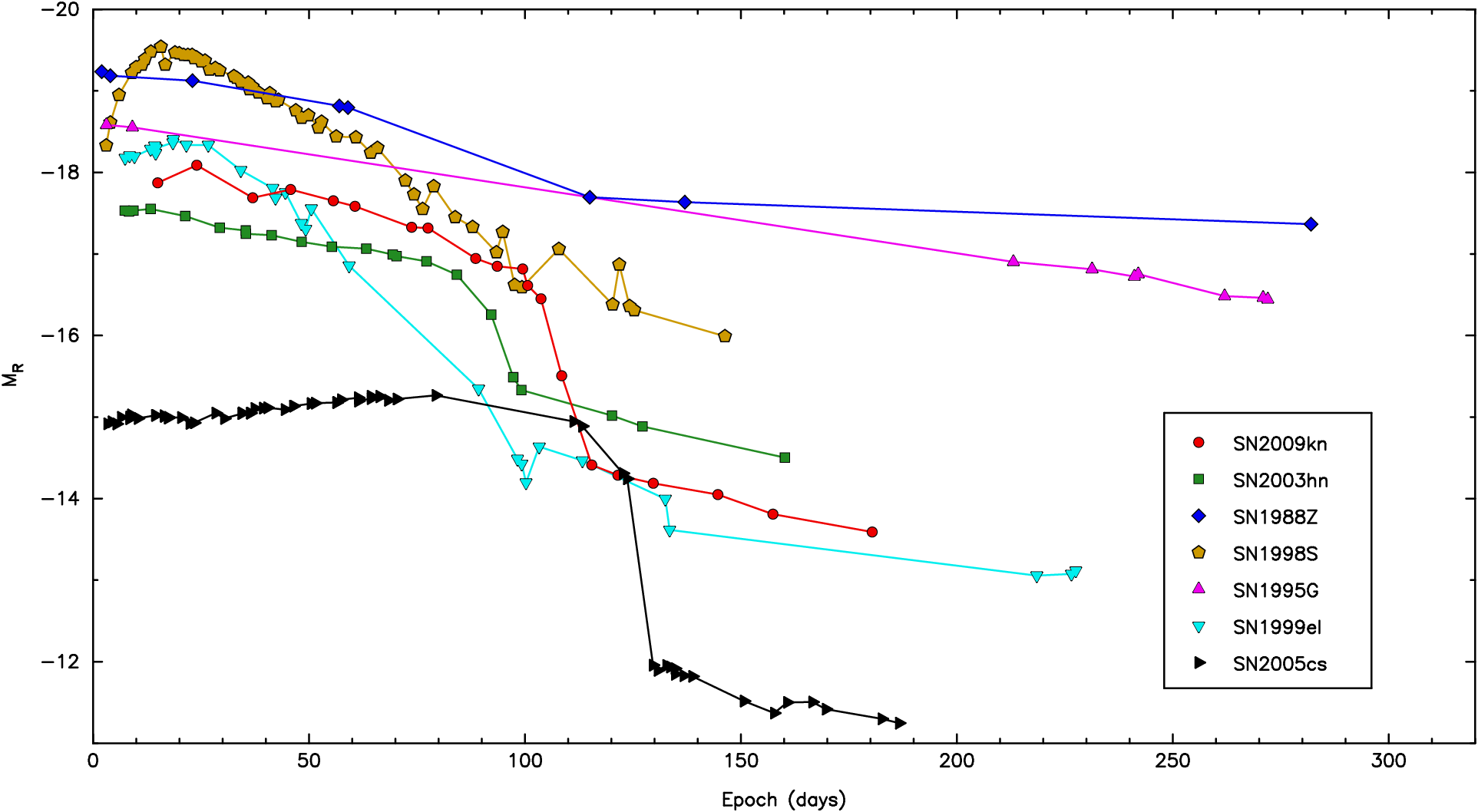}
\caption{Comparison of the absolute \textit{R}-band light curve of SN~2009kn with other SNe. We show the Type IIP SN~2003hn \citep{krisciunas09}, the Type IIn SNe~1988Z \citep{aretxaga99} and 1998S \citep{fassia00,liu00}, two Type IIn events which share spectral similarity to SN~2009kn: SN~1995G \citep{pastorello02} and SN~1999el \citep{dicarlo02}, and the low-luminosity Type IIP SN~2005cs \citep{pastorello09}. Epochs for the SNe are given from the time of explosion where available, otherwise from time of discovery (SNe~1988Z and 1995G).}
\label{r}
\end{figure*}

In Fig.~\ref{r}, we compare the absolute \textit{R}-band light curve of SN~2009kn with the light curves of a number of other SNe. The plateau-like decline, the drop at epoch 100$-$120~d and the radioactive $^{56}$Co tail observed for SN~2009kn are very similar to those of a Type IIP SN, e.g. SN~2003hn, although other evidence points to a different powering mechanism. Compared to other spectroscopically classified Type IIn SNe (e.g. the well-followed SN~1998S), SN~2009kn is very different. It shows no sign of the fast decline at early times shown by SNe 1998S and 1999el, whose light curves more closely resemble Type IIL SNe than a Type IIP. The photometric evolution of SN~2009kn is also very much unlike some spectroscopically similar Type IIn events, e.g. SN~1995G.

\begin{figure}
\begin{minipage}{85mm}
\includegraphics[width=85mm, clip]
{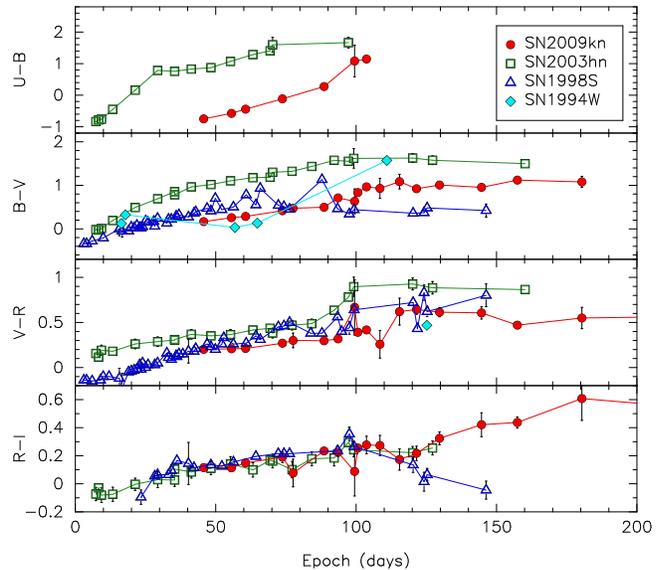}
\caption{Colour evolution of SN~2009kn compared to other SNe $-$ a Type IIP SN~2003hn \citep{krisciunas09} and a Type IIn SN~1998S \citep{fassia00,liu00}. The few available uninterpolated colour values for SN~1994W are shown for completeness. All colour points are reddening corrected}
\label{colour}
\end{minipage}
\end{figure}

The colour curves of SN~2009kn are shown in Fig.~\ref{colour}, together with the colour evolution of some other SNe. The colour evolution of SN~2009kn is fairly similar to that of SN~1998S, whereas SN~2009kn is much bluer than the Type IIP SN~2003hn in all the optical colours. This is particularly apparent in \textit{U$-$B} at the early epochs. However, the \textit{U$-$B} colour of SN~2009kn shows a significant evolution and increases from $-$0.8~mag to 1.1~mag from 46 to 104~d. On day 46, the \textit{B$-$V}, \textit{V$-$R}, and \textit{R$-$I} colours are all $\sim$0.2~mag and increase to 1.0, 0.4, and 0.3~mag, respectively, by day $\sim$100 when the plateau phase ends. During the tail phase, the \textit{B$-$V} and \textit{V$-$R} colours do not change, while \textit{R$-$I} increases to $\sim$0.6~mag by day 180.  

Taking into account the overall behaviour of the light curves, SN~2009kn bears a closer resemblance to a Type IIP SN than to SN~1998S. However, the colour curves are bluer than for a Type IIP and similar to those of SN~1998S. The drop after the plateau phase for SN~2009kn (and SN~1994W) is not particularly large compared to Type IIP SNe. \citet{elmhamdi03} presented a sample of Type IIP SNe that showed a drop in brightness in the range of 1.5$-$3~mag. Even larger drops have been observed for subluminous Type IIP SNe. SN~2005cs dropped by $\sim$3.8~mag in \textit{V} during a period of three weeks after the plateau phase \citep{pastorello09}. We also note that a fairly luminous Type IIP SN~2007od dropped $\sim$6~mag from the plateau to the first observed tail phase slope in roughly 200~d \citep{inserra11}.

\subsection{Bolometric light curves}

Pseudo-bolometric \textit{UBVRI}, \textit{JHK} and interpolated \textit{UBVRIJHK} light curves were created from available photometry of the SNe included in the comparison. The extinction-corrected magnitudes were converted into fluxes and integrated over the filter range using Simpson's rule and the integrated fluxes were converted to luminosities taking into account the distance moduli. 

\begin{figure}
\begin{minipage}{85mm}
\includegraphics[width=85mm, clip]
{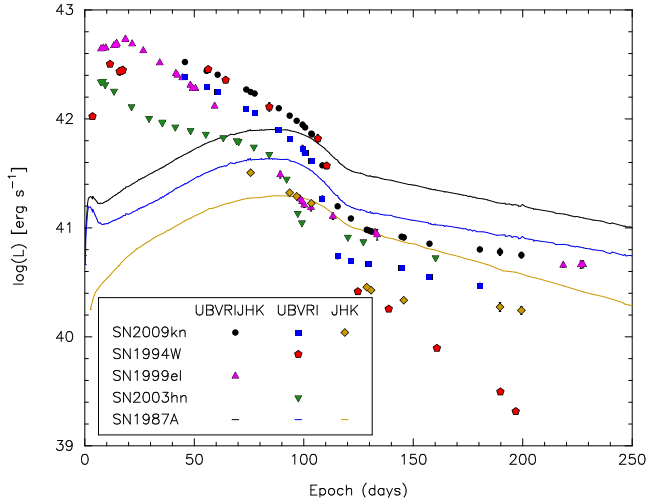}
\caption{Pseudo-bolometric light curves of SN~2009kn compared to those of SN~1987A, SN~1994W, SN~1999el and SN~2003hn.}
\label{bol}
\end{minipage}
\end{figure}

Both the well-sampled optical data of SN~2009kn and the less comprehensive NIR data can be extrapolated and interpolated to all the epochs of observations to derive a \textit{UBVRIJHK} pseudo-bolometric light curve. This is compared to those of several other SNe in Fig.~\ref{bol}. When comparing with SN~1987A, we note that in the early tail phase the pseudo-bolometric light curve follows the radioactive decay fairly closely. Using least-squares fitting between days 129 and 199, the best fit for the light-curve slope is $\gamma=0.90\pm0.06$~mag (100 d)$^{-1}$. This is slightly slower than the pure radioactive value, and could be explained by ongoing CSM interaction in the tail phase. Epochs beyond day 199 were excluded from the tail phase fit because of limited photometric coverage. Since spectral observations seem to indicate that there is still CSM interaction ongoing in the early tail phase, the $^{56}$Ni mass we can derive from the light curve slope should be considered as an upper limit. Assuming complete $\gamma$-ray trapping in the tail phase, the luminosity and $^{56}$Ni mass ratio of SN~2009kn can be derived by comparison with SN~1987A. Comparing the pseudo-bolometric \textit{UBVRIJHK} luminosities of SNe 2009kn and 1987A at day 146 and taking a $^{56}$Ni mass of 0.069~M$_{\odot}$ for SN~1987A \citep{bouchet91}, we estimate a $^{56}$Ni mass $\oldleq$ 0.023 M$_\odot$ for SN~2009kn. As a comparison, SCL98 derived a $^{56}$Ni mass $\oldleq$ 0.015 M$_\odot$ for SN~1994W. No sign of a NIR excess is seen in the \textit{JHK} pseudo-bolometric light curve of SN~2009kn during the first 200~d. 

\subsection{NIR evolution}

Up to day $\sim$200, the NIR evolution of SN~2009kn broadly follows that of the other optical bands (see Fig.~\ref{lc}). Between days 200 and 400, there is no NIR coverage; nevertheless, beyond day 400, the light curves in $H$ and $K$ are relatively flat, while the optical emission continues to drop steeply. This behaviour is suggestive of dust condensation.

To examine this possibility, we first fitted a single blackbody to the NIR spectral energy distribution (SED) at day 417 for which date photometry in \textit{J}, \textit{H} and \textit{K} bands was available. To avoid the effects of possible contamination from line emission in the \textit{J} band, we also carried out the fit using only the \textit{H}- and \textit{K}-band points. These fits all yielded similar blackbody temperatures of $\sim$1400$-$1500~K and radii of $\sim$2.6$-$3.0$ \times 10^{15}$~cm. The blackbody temperature is consistent with thermal emission from dust. The blackbody radius is consistent with an expansion velocity of $\sim$800~km~s$^{-1}$ (according to the P Cygni minima velocity during the plateau phase; see Section~4.5) assuming a small initial radius. Thus, the IR emission could be seen to originate within the same dense shell that gave rise to the optical emission from the SN. This behaviour is in line with other type IIn SNe. For SN~1998S, a strong late-time NIR excess was also observed and was at least partly attributed to newly formed dust in a cool dense shell (CDS) formed behind the reverse shock \citep{pozzo04}. The $H-K$ colour of $+1.2$ at day 417 is comparable to that of SN~1998S ($+1.5$ at day 333). \citet{meikle06} derived a dust evaporation radius of $\sim$3$ \times 10^{16}$~cm for the type IIP SN~2002hh, appropriate for graphite grains (for silicate grains this radius would be larger). However, SN~2009kn was $\sim$1~mag brighter at early times than SN~2002hh \citep{pozzo06}, which implies a larger dust evaporation radius. The dust evaporation radius of SN~2009kn is therefore at least an order of magnitude larger than the derived warm dust blackbody radius of SN~2009kn. This also supports the idea that dust formed between days 200 and 400, since the time-scales are consistent with those observed previously \citep[see e.g.][]{kotak09,meikle11}.

Unfortunately, no spectroscopy of SN~2009kn is available beyond day 225 to confirm that new dust indeed formed. We compared the day 120 spectrum with the day 225 spectrum to search for blueshifted lines which would further support our claim. However, we do not detect any noticeable blueshift of the H$\alpha$ emission peak; if anything, the line peak shifted slightly to the red. Our last $K$-band spectrum (day 225) does not show a strongly rising continuum, and there is no sign of molecular emission due to CO -- a commonly observed precursor to dust formation \citep[e.g.][]{kotak05}. Again, this is not inconsistent with the time-scales and $HK$ evolution described above. SN~1987A showed clear signatures of dust formation only as late as day 600  \citep{wooden93}, although it did show molecular emission at earlier epochs.

No NIR data are available for SN~1994W. Assuming that SN~1994W was a genuine SN, then partial trapping of $\gamma$-rays in the tail phase or dust formation already at $\sim$100~d could explain its steep $R$-band evolution during the tail phase, leading SCL98 to underestimate the amount of $^{56}$Ni that it produced.

\section{Spectroscopy}

Spectroscopic observations of SN~2009kn were obtained at 13 epochs from day 35 to 225 post-explosion. NOT/ALFOSC, NTT/EFOSC2 and the 8.2-m Very Large Telescope (VLT) with Ultraviolet and Visual Echelle Spectrograph \citep[UVES;][]{dekker00} and the optical/NIR spectrograph X-Shooter \citep{vernet11} were used to obtain optical spectra. The NIR spectroscopy was obtained with X-Shooter [simultaneously with ultraviolet (UV) and optical], NTT/SofI and LBT/LUCIFER. 

\subsection{Data reduction}

The data reduction for the NOT and NTT spectra were done using standard {\sc iraf} tasks and the {\sc quba} pipeline. The raw data were bias subtracted and flat-field corrected. Spectroscopic flats were obtained before and after the target exposure with the same pointing as for SN~2009kn as necessary to minimize the possible fringing effects. The spectra were extracted with the {\sc iraf} task {\sc apall}, wavelength calibrated using arc lamp exposures and cross-correlated with the sky lines. The second-order contamination for grism \#4 at the NOT (20 per cent of the \textit{U}-band flux and 5 per cent of the \textit{B}-band flux visible beyond 5900 \AA) was corrected using the method presented by \citet{stanishev07}. The LBT spectrum was extracted, reduced and calibrated using the LUCIFER pipeline. The relative flux calibration of the optical spectra was done by deriving the sensitivity curve for the instrument set-up from spectroscopic standard star observations. The NIR spectra were calibrated by dividing the extracted spectrum with a spectrum of a telluric standard star observed at a similar airmass and close in time as the SN and multiplied with the same spectral type stellar library spectrum \citep{pickles98}. 

UVES spectrum was pipeline reduced using the Reflex-based workflow provided by ESO. X-Shooter spectra were pre-reduced using version 1.1.0 of the dedicated ESO pipeline \citep{goldoni06}, with calibration frames (biases, darks, arc lamps and flatfields) taken during daytime. The 1D spectrum extractions were done using {\sc apall} in a similar way to the low-resolution spectra. Relative flux calibrations were performed through spectrophotometric standards taken from the ESO list,\footnote{http://www.eso.org/sci/facilities/paranal/instruments/ xshooter/tools/specphot\_list.html} which includes stars with well-known emission fluxes from UV to NIR. We removed the telluric bands using telluric standards taken soon after each scientific exposures and at the same airmass as the SN.

All the spectra were absolute flux calibrated with \textit{BVRIJHK} broadband photometry interpolated/extrapolated to the observation epoch. An average scaling factor was derived for each epoch and multiplied to the spectrum over the whole wavelength range. Finally, all the spectra were corrected to the heliocentric reference frame. For a complete log of spectroscopic observations, see Table~4.

\begin{table*}
\begin{center}
\caption{Spectroscopy of SN~2009kn}
\begin{tabular}{cccccccc}
\hline
JD & Epoch & Grism & Range & Resolution & PA$^{a}$ & Exp. time & Telescope/instrument\\ 
(2400000+) & (d) & & (\AA) & (\textit{R}) & ($^{\circ}$) & (s) & \\ \hline
55150.7 & 35 & Blue,red & 3300-6800 & $\sim$40000 & 0 & 2300 & VLT/UVES \\
55161.7 & 46 & gm\#4 & 3200-9100 & 270 & 12 & 900 & NOT/ALFOSC \\
55171.6 & 56 & gm\#7 & 3850-6850 & 650 & 157 & 900 & NOT/ALFOSC \\ 
55177.8 & 62 & UVB,VIS,NIR & 3000-24800 & 5100,8800,5100 & 47 & 4x150,4x150,4x3x50 & VLT/X-Shooter \\ 
55189.7 & 74 & gm\#11,gm\#16 & 3380-10320 & 340,510 & 60 & 1800,1800 & NTT/EFOSC2 \\ 
55191.8 & 76 & BG,RG & 9500-25200 & 600,600 & 0 & 6x3x180,6x3x180 & NTT/SofI \\ 
55212.7 & 97 & UVB,VIS,NIR & 3000-24800 & 5100,8800,5100 & 70 & 4x150,4x150,4x3x50 & VLT/X-Shooter \\ 
55212.9 & 97 & 200H+K & 15000-25000 & 1900,2600 & 51 & 24x300 & LBT/LUCIFER \\ 
55215.5 & 100 & gm\#4 & 3200-9100 & 270 & 148 & 3x1000 & NOT/ALFOSC \\ 
55231.6 & 116 & gm\#4 & 3200-9100 & 270 & 27 & 3x1000 & NOT/ALFOSC \\ 
55235.6 & 120 & UVB,VIS,NIR & 3000-24800 & 5100,8800,5100 & 71 & 4x650,4x650,4x3x270 & VLT/X-Shooter \\
55245.7 & 130 & gm\#16 & 6015-10320 & 510 & 80 & 2x3600 & NTT/EFOSC2 \\ 
55340.5 & 225 & UVB,VIS,NIR & 3000-24800 & 5100,8800,5100 & 107 & 4x650,4x650,4x3x270 & VLT/X-Shooter \\
\hline
\end{tabular}
\end{center}
\begin{flushleft}
$^{a}$ The position angle (PA) of the slit on the sky is measured from north to east.
\end{flushleft}
\end{table*}

\subsection{Epoch of explosion and extinction}

For the Galactic line-of-sight extinction, we used the value derived by \citet{schlegel98} dust maps with a reddening of $E(B-V)=0.114$~mag. We conclude from the early epoch spectra that the host galaxy extinction is negligible compared to the Galaxy due to the absence of redshifted \NaID\ $\lambda\lambda$~5889.95,~5895.92 lines in our data, whilst the rest wavelength Galactic lines are clearly identified [see \citet{turatto03} for details but also \citet{poznanski11} for a caveat on the use of \NaI\ as a reddening indicator]. Thus, using the \citet{cardelli89} reddening law with $R_{V}=3.1$, we adopt a total line-of-sight extinction of $A_V=0.353$~mag (i.e. only the Galactic reddening). 

Due to the lack of observations of the SN field close in time before the discovery, we have no independent estimate of the explosion date beyond the discovery data. We take instead the simple approach of comparing SN~2009kn with SN~1994W, which has a well-defined explosion date of JD~$=2449548_{-4}^{+2}$ (SCL98). We used the SuperNova IDentification ({\sc snid}) code \citep{blondin07} to cross-correlate first five optical spectra of SN~2009kn (see Table~4) against the spectra of SN~1994W presented by CBC04, along with the library of spectra included in {\sc snid} (version 5.0 with template set 2.0). For the spectra included in the analysis, {\sc snid} always returned spectra of SN~1994W as the best two matches. Other suggestions were mainly luminous blue variables (LBVs), active galactic nuclei, and other non-SN sources. We used the rlap value output of {\sc snid} as an estimate of the quality of the best fits and calculated a weighted average for the explosion date of 15~d before the discovery. The standard deviation of the explosion dates obtained using different spectra, summed in quadrature with the error derived for the explosion date of SN~1994W, was adopted as the final error. We thus adopt an explosion date of JD~$=2455116_{-11}^{+10}$, noting that it gives a very similar light curve drop epoch for SN~2009kn as for SN~1994W. The applied method assumes that the two SNe are absolutely identical. While this may not be the case, comparison with the spectral development of other SNe (see Section~5.2) supports its use for estimating the explosion date.

\subsection{Spectral evolution}

\begin{figure*}
\includegraphics[height=130mm, clip]
{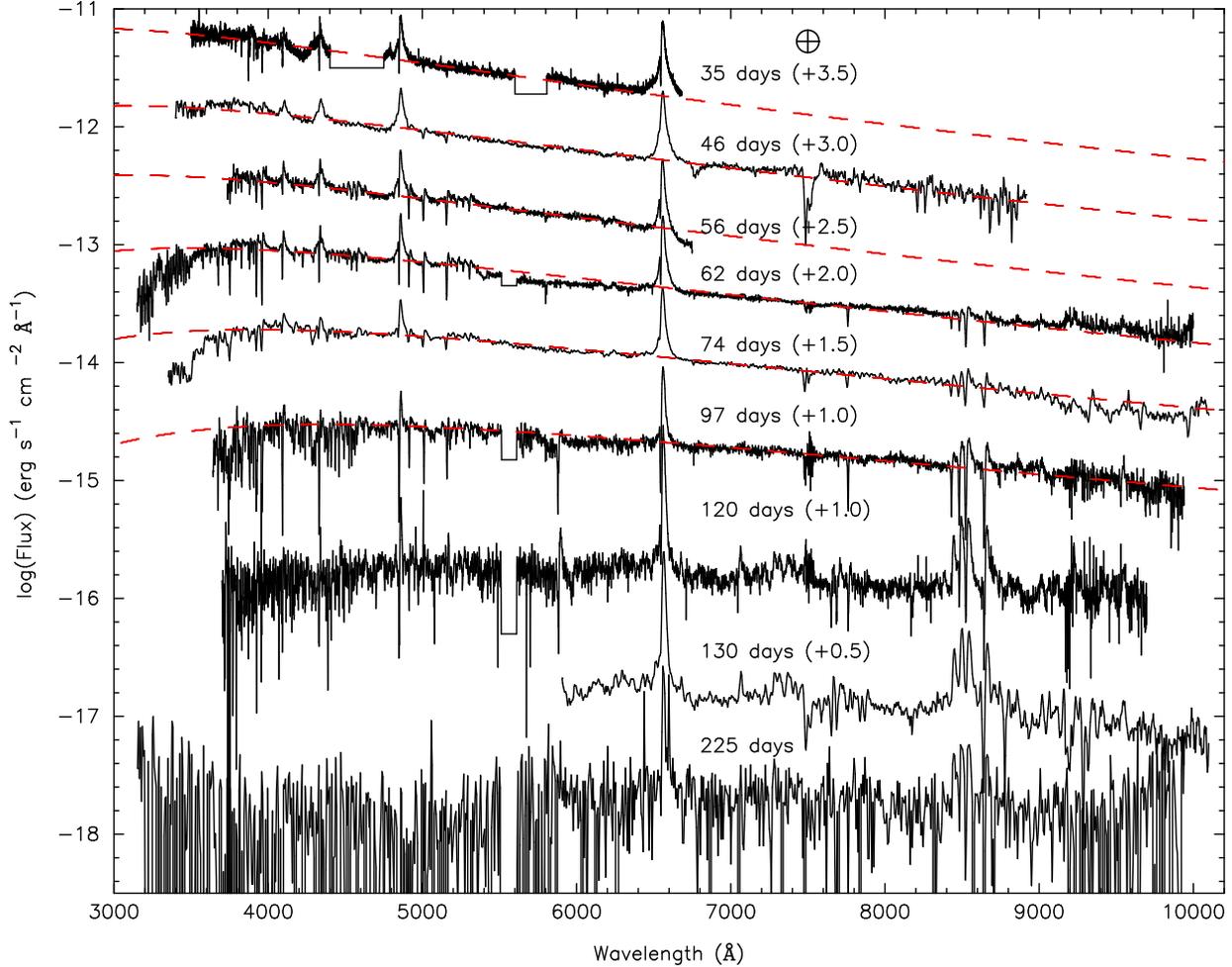}
\caption{Optical spectral evolution of SN~2009kn. The spectra are dereddened and the wavelengths corrected to the host galaxy rest frame. The blackbodies fitted to the early spectra have been overplotted with the red dashed curves (see Section~4.4 for details). Atmospheric absorption features have mostly not been removed. The wavelength of the most significant telluric feature is indicated with a $\oplus$ symbol. The spectra have been vertically shifted for clarity as indicated in the parentheses.}
\label{spect}
\end{figure*}

\begin{figure*}
\includegraphics[height=130mm, clip]
{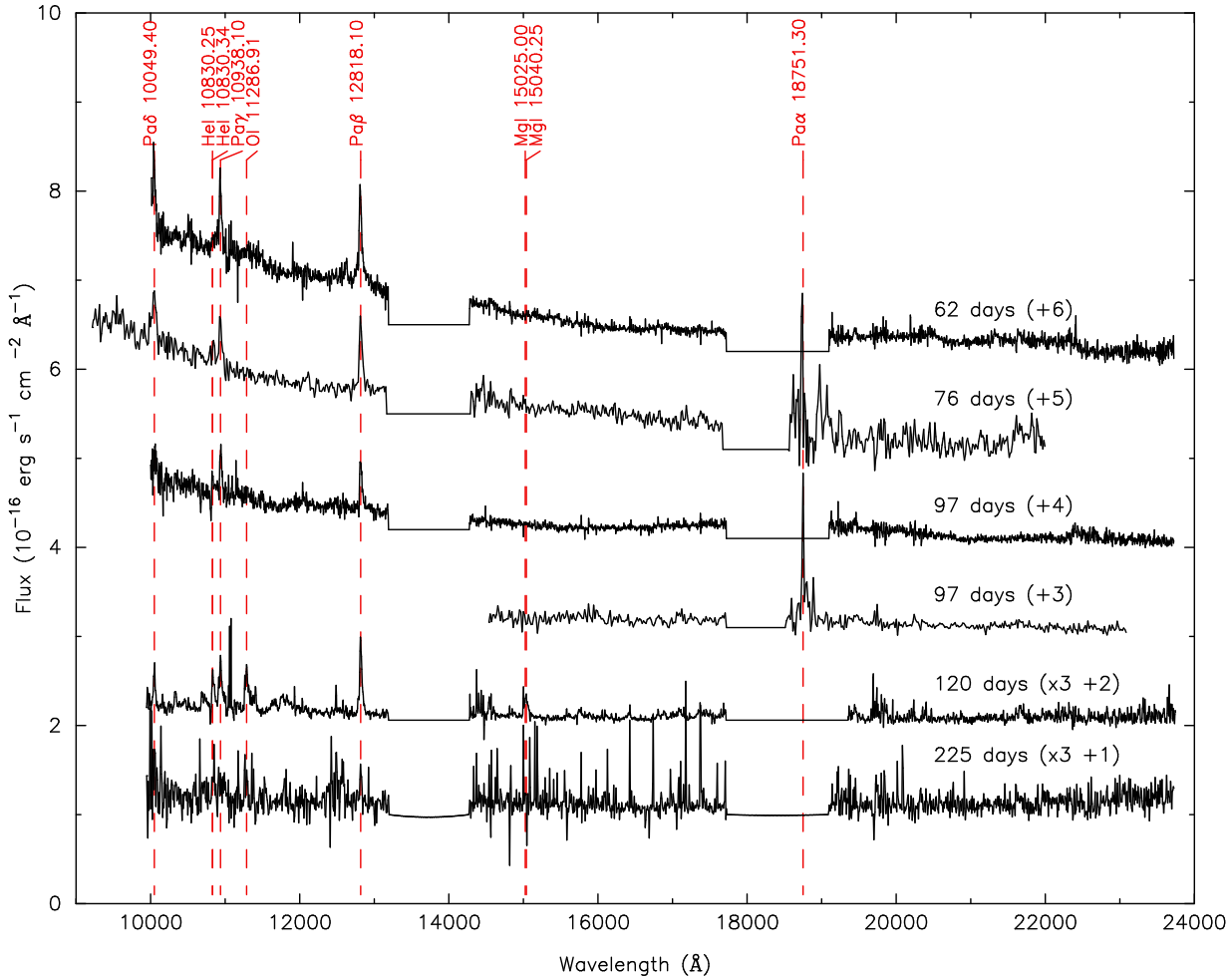}
\caption{NIR spectral evolution of SN~2009kn. The spectra are dereddened and the wavelengths corrected to the host galaxy rest frame. Atmospheric absorption features have been cut from the spectra. The spectra have been vertically shifted as indicated in the parentheses.}
\label{spect2}
\end{figure*}

The spectral time series of SN~2009kn in the optical is shown in Fig.~\ref{spect}. The NIR spectral evolution is shown in Fig.~\ref{spect2} with the identified NIR lines overlaid. The day 62 UV-optical spectrum of SN~2009kn has already been presented in a compilation of recent observations of Type IIn SNe with X-Shooter by \citet{pastorello11}. Following the line identifications carried out for SN~1994W initially by CBC04 and later very comprehensively by DHG09 (see also the supplementary online data in their appendix A), we performed our own line identifications for SN~2009kn to allow more detailed comparison between the two events. 

For analytical line identification purposes all the 1D spectra were extracted from the 2D frames using the {\sc iraf} task {\sc apall} with optimal extraction enabled, producing a target spectrum and also a weight spectrum describing the pixel-to-pixel noise level. Due to the higher resolution of the X-Shooter spectra, the 2D spectra were binned before the extraction to enable {\sc apall} to produce a more reliable noise spectrum. We took care that even the narrowest features were not lost in binning the spectra and that different epochs were binned similarly. A continuum level for the 1D spectrum was determined by heavily median smoothing the spectrum with a 100 \AA\ window. Only features that differed by more than three times the wavelength-dependent noise level from the continuum were accepted as real lines, i.e. emission and absorption features with $\gtrsim3\sigma$ peak detections. Further rejection of features was done by inspecting closely the 2D frames to identify false detections caused by fringing effects, cosmic rays, telluric lines or failed tracing and extraction in the end regions of the spectra with intrinsically low signal-to-noise ratio. For example, no lines were identified in the \textit{K}-band region of the spectra. In the low resolution spectra, only the lines that were identified also in at least one of the higher resolution X-Shooter spectra were accepted as real. We note that our method is more conservative than the approach used by CBC04. To identify most of the atomic lines, we used the National Institute of Standards and Technology (NIST) Atomic Spectral Database \citep{ralchenko08}, the line tables of CBC04 and DHG09, and the \FeII\ line tables of \citet{sigut03}.

To assist in the line identification in the photospheric phase optical spectra, we also used a Monte Carlo atmosphere code similar to that of \citet{mazzali93} and the atomic line list from \citet{jerkstrand11}. The code treats electron and line scattering in the nebular non-local thermodynamic equilibrium approximation given density, abundances, a thermalization radius and the observed luminosity. The temperature at the thermalization radius is iterated to reproduce the observed luminosity. The values for these quantities were taken from DHG09 for the day 79 spectrum of SN~1994W. Except for the hydrogen lines, which are mainly recombination driven, the lines in the day 79 spectrum of SN~1994W are well reproduced by this procedure. Given the strong similarity between SN~1994W and SN~2009kn in the photospheric phase, we assume these identifications to apply to SN~2009kn as well. For complete lists of identified lines, see Tables~A1 and A2 in the Appendix A.

\begin{figure}
\begin{minipage}{85mm}
\includegraphics[width=85mm, clip]
{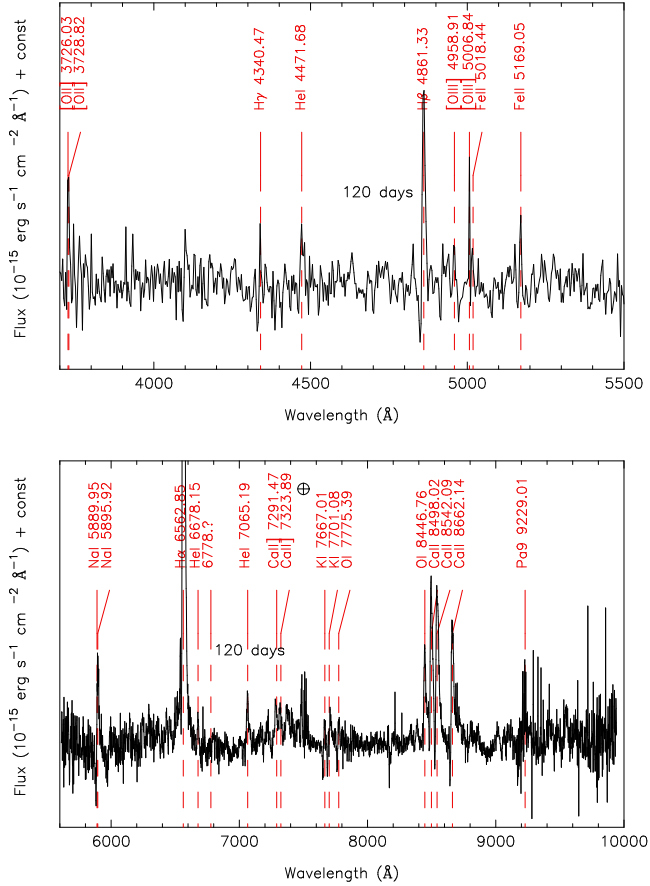}
\caption{Day 120 UVB and VIS arm X-Shooter spectrum with the identified lines overlaid. The spectra are dereddened and the wavelengths corrected to the host galaxy rest frame.}
\label{xshoot2}
\end{minipage}
\end{figure}

In Figs.~\ref{xshoot} and \ref{xshoot2}, the lines identified in the X-Shooter spectra are shown. The medium-resolution spectra from X-Shooter enabled us to fully resolve the narrow absorption components of multiple lines. These spectra show the main spectral evolution points of SN~2009kn in an otherwise only slowly changing time series. 

\begin{figure}
\begin{minipage}{85mm}
\includegraphics[width=85mm, clip]
{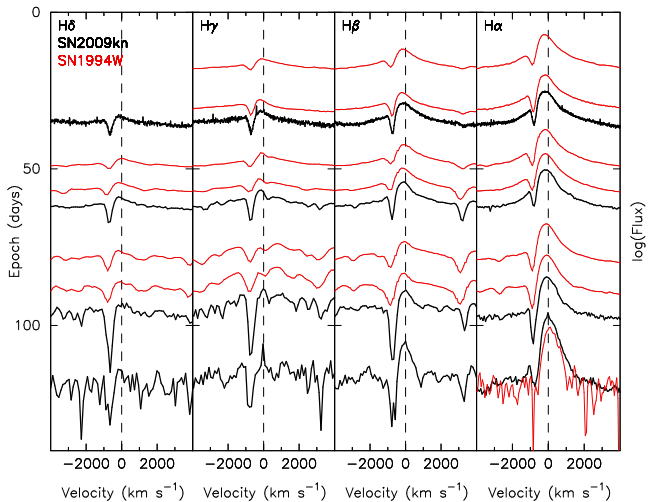}
\caption{Evolution of Balmer line profiles of SN~2009kn compared with SN~1994W. Spectra are shown in logarithmic scale vertically shifted for clarity to match with the estimated epoch.}
\label{bal}
\end{minipage}
\end{figure}

The spectrum of SN~2009kn is dominated by strong Balmer lines, especially in the early epochs, and these lines show similar evolution to that of SN~1994W (for the comparison, see Fig.~\ref{bal}). The slightly less prominent absorption components of SN~1994W compared to SN~2009kn can be explained by the poorer spectral resolution in the SN~1994W data. We also note the similarity in the wavelength evolution of the Balmer line peaks, gradually shifting redwards as a function of time. The flux measurements of the hydrogen lines are reported in Table~5. The H$\beta$ line flux was corrected for blending with \FeII\ $\lambda$~4923.94. H$\gamma$ is also affected by multiple faint \FeII\ lines but these effects were not corrected for. The H$\alpha$ flux on day 56 may be affected by poor flux calibration as the line lands in the blue end of the grism. The early reddening-corrected Balmer decrement H$\alpha$:H$\beta$:H$\gamma = 1.3:1:0.5$ on day 46 is very similar to the reported value of $1.5:1:0.6$ by CBC04 for SN~1994W on days 46 and 56. By day 62, the Balmer decrement of SN~2009kn evolved to H$\alpha$:H$\beta$:H$\gamma = 1.5:1:0.2$. During the late plateau phase, H$\gamma$ can no longer be measured properly. However, H$\alpha$/H$\beta$ has a value of 2.1 by day 74 and 7 by day 97, similar to SN~1994W with H$\alpha$/H$\beta=2.7$ by day 79 and 5.2 by day 89. In the day 120 tail phase spectrum, the ratio has increased to H$\alpha$/H$\beta=16$ for SN~2009kn. The evolution of the Balmer decrement can be qualitatively understood as a result of transition of the recombination spectrum from a regime of optically thick, almost thermalized Balmer lines towards a regime of optically thick Balmer lines but still far from complete thermalization.

\begin{table}
\begin{center}
\caption{Reddening-corrected hydrogen line fluxes in units of $10^{-15}$~erg~s$^{-1}$~cm$^{-2}$.}
\begin{tabular}{cccccccc}
\hline
Epoch (d) & H$\alpha$ & H$\beta$ & H$\gamma$ & Pa$\beta$ & Pa$\gamma$ & Pa$\delta$\\ \hline
46 & 66 & 51 & 27 & - & - & - \\
56 & 45 & 40 & 15 & - & - & - \\
62 & 45 & 29 & 5.6 & 6.4 & 3.2 & - \\
74 & 35 & 17 & - & - & - & - \\
76 & - & - & - & 4.4 & 2.7 & 3.1 \\
97 & 17 & 2.4 & - & 2.1 & 1.6 & - \\
120 & 5.3 & 0.32 & - & 1.2 & 0.85 & 0.55 \\
\hline
\end{tabular}
\end{center}
\end{table}

In the NIR, the Paschen lines Pa$\beta$ and Pa$\gamma$ are the most prominent features. Pa$\alpha$ overlaps with the strong telluric band between \textit{H} and \textit{K}. Evidence of the line can be seen in the day 76 NTT spectrum and in the day 97 LBT spectrum. The Pa$\beta$/Pa$\gamma$ ratio decreases slowly from $\sim$2 at day 62 to $\sim1.3-1.4$ both before and after the light curve drop at day 97 and day 120. 

\begin{figure}
\begin{minipage}{85mm}
\includegraphics[width=85mm, clip]
{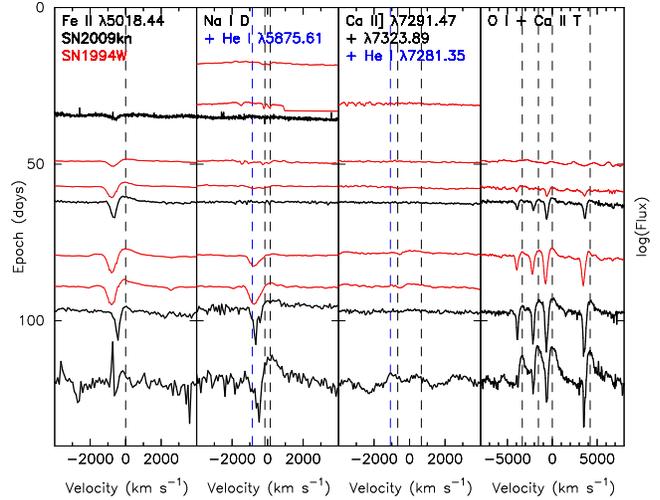}
\caption{Evolution of an example \FeII\ line at $\lambda$~5018.44, \NaID, weak semiforbidden \CaII] $\lambda\lambda$~7291.47,~7323.89, and the \CaII\ NIR triplet of SN~2009kn compared with SN~1994W. The \HeI\ lines at $\lambda$~5018.44 and $\lambda$~7281.35 are potentially blended with other lines marked in the middle panels. Scaling as in Fig.~\ref{bal}.}
\label{line}
\end{minipage}
\end{figure}

No optical \HeI\ lines were detected before the light curve drop, including the broad $\lambda$~5875.61 visible in the early spectra of SN~1994W and SN~2005cl \citep[see CBC04 and][respectively]{kiewe12}. In the tail phase, emission lines with a good match to \HeI\ $\lambda\lambda$~4471.68,~6678.15,~7065.19 emerge. There is no clear detection of \HeI\ at $\lambda$~5875.61. We believe that the strong P Cygni line associated with \NaID\ is preventing us from properly observing the underlying feature of the \HeI\ $\lambda$~5875.61 line. We associate the strong P Cygni line itself with \NaID\ rather than \HeI\ due to its convincing rest wavelength match to \NaID (see Fig.~\ref{line}). A \HeI\ $\lambda$~5875.61 line with flux similar to the identified \HeI\ $\lambda$~7065.19 line could be blended with a strong absorption of the \NaID\ lines in the day 120 spectrum remaining undetected in the observed spectrum. Similarly, the \HeI\ $\lambda$~5015.68 line may be blended with \FeII\ $\lambda$~5018.44 and the \HeI\ $\lambda$~7281.35 line with the \CaII] $\lambda$~7291.47. Other optical \HeI\ lines such as $\lambda\lambda$~3888.64,~4026.19 remain absent. CBC04 did not report any detection of \HeI\ lines at late times, although in the tail phase spectra this can be explained by a low signal-to-noise ratio, limited wavelength coverage or line blending. In the NIR, the day 97 and day 120 X-Shooter spectra of SN~2009kn show an \HeI\ P Cygni line at $\lambda$~10830 before and after the drop; however, there is no sign of $\lambda$~20581 \HeI\ line in any of the NIR spectra.

The epochs at which different lines emerge and disappear are very similar between SN~1994W and SN~2009kn. For the evolution of a few selected metal line profiles of SN~2009kn compared to the time series of SN~1994W, see Fig.~\ref{line}. As with the Balmer series lines, this comparison demonstrates the striking similarity in the spectral evolution between SN~1994W and SN~2009kn. 

The spectra of SN~2009kn contain numerous Fe lines, many of which, such as \FeII\ $\lambda\lambda$~4233.12,~4923.94,~5018.44, are already above the detection limit in the early spectra on day 46 or day 56 and stay prominent throughout the plateau phase. Most of the \FeII\ lines exhibit a clear P Cygni component with the absorption dominating over the emission component. After the light-curve drop, the \FeII\ lines essentially disappear. 

The most crowded regions of lines in the optical contain not only a forest of \FeII\ lines but also multiple lines which we associate with \TiII, such as $\lambda\lambda$~4395.03,~4443.79. These were also identified by DHG09, and compared to CBC04 this is the major difference between our line identifications. \TiII\ lines are no longer detectable after the drop.

Similar to many of the \FeII\ lines, \CaII\ H\&K $\lambda\lambda$~3933.66,~3968.47 (latter blended with H$\epsilon$) emerge on day 56 and remain through the plateau phase with the emission component fading over time leaving absorption-dominated profiles on day 97. The evolution of the \CaII\ triplet $\lambda\lambda$~8498.02,~8542.09,~8662.14 is completely opposite to \CaII\ H\&K. In the first X-Shooter spectrum on day 62, the \CaII\ triplet is present with strong absorption components. Over time, the \CaII\ triplet becomes more prominent with a clear emission component starting to dominate the line profiles. After the light-curve drop, the \CaII\ triplet lines are the most prominent after H$\alpha$.

The \OI\ $\lambda$~8446.76 line emerges in the day 62 X-Shooter spectrum together with the \CaII\ NIR triplet and \OI\ $\lambda$~7775.39. Both \OI\ lines are still visible in the tail phase spectra. The tail phase NIR spectrum on day 120 also shows the \OI\ $\lambda$~11286.91 line appearing, presumably pumped by Ly$\beta$. 

\NaID\ emerges in the second X-Shooter spectrum on day 97. No clear sign of the \NaID\ is seen in the low resolution day 74 NTT spectrum. For comparison, \NaID\ emerged with a clear P Cygni profile in the day 76 spectrum of SN~1994W. After the drop, \NaID\ is one of the most prominent lines. 

Among the other new features in the two tail phase spectra on day 120 and 130, two new P Cygni lines are detected which we associate with \KI\ $\lambda\lambda$~7667.01,~7701.08. These lines were also observed in the 2009 optical transient in UGC 2773 \citep{smith10b}, which showed spectroscopic similarities to SN~2009kn.

In our day 120 X-Shooter multiple, very narrow forbidden lines emerge which are not associated with the SN: the clear [\OII] doublet $\lambda\lambda$~3726.03,~3728.82, unresolved [\OIII] at $\lambda\lambda$~4958.91,~5006.84 and \CaII] $\lambda\lambda$~7291.47,~7323.89, as well as narrow Balmer line components. For the [\OII] lines we measure a $V_{\mathrm{FWHM}}=90$~km~s$^{-1}$ and for H$\alpha$ $V_{\mathrm{FWHM}}=60$~km~s$^{-1}$. By measuring the [\OII] line fluxes, we estimate a [\OII] $\lambda$~3728.82/$\lambda$~3726.03 ratio of $1.19\pm0.17$, typical of \HII\ regions \citep{osterbrock06}. Using the radiative transition probabilities $A_{^{4}S_{3/2}-^{2}D_{3/2}}=1.59 \times 10^{-4}$ and $A_{^{4}S_{3/2}-^{2}D_{5/2}}=2.86 \times 10^{-5}$ from NIST and collision strengths $\Omega(^{4}S_{3/2}-^{2}D_{3/2})=0.585$, $\Omega(^{4}S_{3/2}-^{2}D_{5/2})=0.883$ and $\Omega(^{2}D_{5/2}-^{2}D_{3/2})=1.426$ from \citet{pradhan06} at 10000~K and using equations (2) and (4) from \citet{seaton54}, we derived a temperature and electron density relation for the [\OII] region. Assuming the temperature to range between 5000 and 20000~K, an electron density of 100$-$200~cm$^{-3}$ was obtained, consistent with an \HII\ region. In the 2D spectral frames, we see an extended emission region superimposed on the position of the SN. Unsuccessful removal during the sky subtraction process of the spectra can explain this without the need to associate the emission with the SN itself. The nearby \HII\ region mentioned in Section~3.1 is located roughly 1.2~arcsec from SN~2009kn at a position angle $\sim$310$^{\circ}$, measured from north to east, corresponding to a slit angle of 130$^{\circ}$. As shown in Table~4, the slit positions of most of the spectra are very different to this. In particular, the 0.9~arcsec slit of the day 120 spectrum which shows the very narrow 60~km~s$^{-1}$ H$\alpha$~component for the first time does not pass through the optical centre of the nearby \HII\ region. However, diffuse emission from the same nearby \HII\ region may well extend into the line-of-sight towards the SN. The earlier spectra, even with high enough resolution, are not likely to show this line feature since the SN, with strong continuum, is much brighter than in the tail phase and any background emission is lost in the noise. To explain the linewidth of the 60~km~s$^{-1}$ H$\alpha$~component as thermal broadening, the required temperature would be of the order of $\sim$80000~K, obviously too high of a temperature for an \HII\ region. Therefore, this velocity dispersion reflects macroscopic, rather than thermal, motions of ionized gas according to a standard interpretation of a large linewidth for luminous extragalactic \HII\ regions \citep{shields90}.

\begin{figure}
\begin{minipage}{85mm}
\includegraphics[width=85mm, clip]
{fig9.ps}
\caption{Day 62 and day 97 UVB and VIS arm X-Shooter spectra with the identified lines overlaid. The spectra are dereddened and the wavelengths corrected to the host galaxy rest frame.}
\label{xshoot}
\end{minipage}
\end{figure}

\subsection{Evolution of the spectral energy distribution}

In order to compare the evolution of the SEDs between SN~2009kn and SN~1994W, we performed blackbody fits to the spectra (see Table~6). A continuum was first fitted to the dereddened and redshift-corrected optical spectra. From the continuum of the spectra, we extracted several points at every 100 \AA\ and derived a best fit for a single blackbody by $\chi^{2}$ fitting. During the plateau phase, the blue end of the optical spectra is affected by severe line blanketing, similar to SN~1994W, and no single blackbody can be fitted to the continuum in this part of the spectra. For comparison, we also carried out blackbody fits solely based on the \textit{UBVRI} photometry. To consistently compare SN~2009kn with SN~1994W, we carried out an absolute calibration of the observed spectra of SN~1994W using interpolated and extrapolated \textit{B}- and \textit{V}-band photometry (see Section~4.1) and derived blackbody properties in a similar fashion as for SN~2009kn. We adopted a distance of 25.4~Mpc and total reddening of $E(B-V)=0.17$~mag for SN~1994W, the same as SCL98, CBC04 and DHG09. This distance is close to the Virgo and Great Attractor corrected luminosity distance of 23.9~Mpc \citep{mould00} for the host galaxy NGC 4041. The blackbodies fitted to the early spectra of SN~2009kn are shown overlaid with the spectra in Fig.~\ref{spect}.

\begin{table}
\begin{center}
\caption{Best blackbody matches to the spectroscopy and photometry of SN2009kn.}
\begin{tabular}{ccccc}
\hline
Epoch & $T$ & $R$ & $\log(L)$ & Data\\ 
(d) & (K) & ($10^{14}$~cm) & (erg~s$^{-1}$) & \\ \hline
35 & 10500 & 7.9 & 41.8 & Spect\\
46 & 9750 & 8.3 & 42.6 & Phot\\
46 & 9430 & 8.6 & 42.6 & Spect\\
56 & 8780 & 8.9 & 42.5 & Phot\\
56 & 9320 & 8.0 & 42.5 & Spect\\
61 & 8270 & 9.4 & 42.5 & Phot\\
62 & 8310 & 9.3 & 42.4 & Spect\\
74 & 6910 & 11 & 42.3 & Phot\\
74 & 7420 & 9.9 & 42.3 & Spect\\
78 & 7770 & 8.8 & 42.3 & Phot\\
89 & 6210 & 11 & 42.1 & Phot\\
94 & 6690 & 9.1 & 42.1 & Phot\\
100 & 5210 & 13 & 41.9 & Phot\\
100 & 6350 & 9.8 & 42.0 & Spect\\
101 & 6170 & 9.3 & 42.0 & Phot\\
104 & 4940 & 13 & 41.9 & Phot\\
\hline
\end{tabular}
\end{center}
\end{table} 

\begin{figure}
\begin{minipage}{85mm}
\includegraphics[width=85mm, clip]
{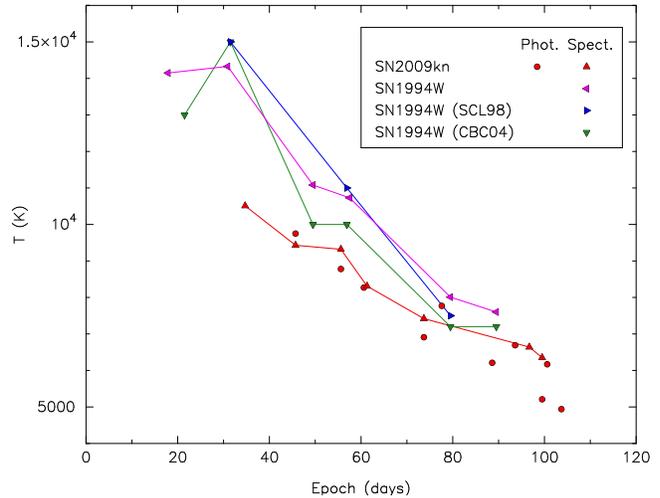}
\caption{The blackbody temperature of SN~2009kn, derived from spectroscopy and photometry. Spectroscopy-based values for SN~1994W from SCL98 and CBC04 are shown for comparison.}
\label{temp}
\end{minipage}
\end{figure}

\begin{figure}
\begin{minipage}{85mm}
\includegraphics[width=85mm, clip]
{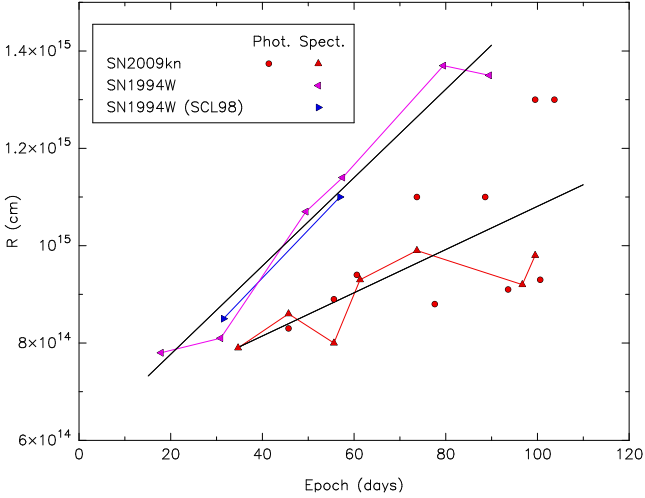}
\caption{Same as Fig.~\ref{temp}, but for the blackbody radius. The black lines indicate slopes fitted to the data of SN~2009kn and SN~1994W}
\label{rad}
\end{minipage}
\end{figure}

We find that the measured blackbody temperatures decrease and radii increase during the plateau phase of SN~2009kn and SN~1994W (see Figs.~\ref{temp} and \ref{rad}). Using least-squares fitting, an average blackbody expansion velocity of $\sim$500~km~s$^{-1}$ was obtained with an initial radius of $6 \times 10^{14}$~cm. In the case of SN~1994W a velocity of $\sim$1000~km~s$^{-1}$ with similar initial radius was found.

\subsection{Line profiles of hydrogen}

\begin{figure}
\begin{minipage}{85mm}
\includegraphics[width=85mm, clip]
{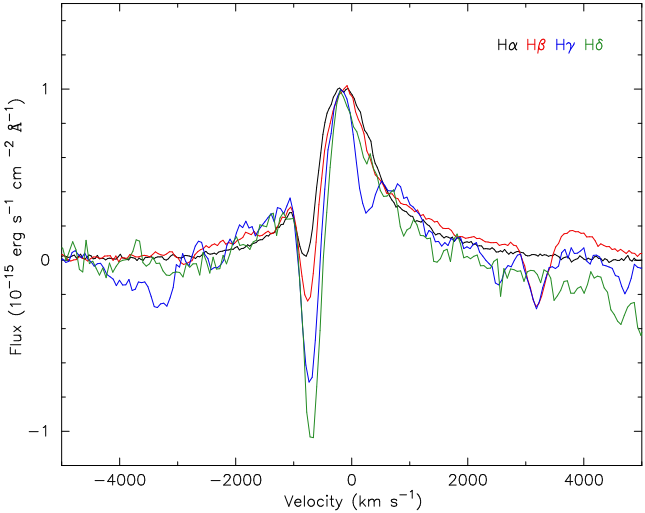}
\caption{The continuum-subtracted and peak-normalised Balmer series line profiles overplotted from the first X-Shooter spectrum on day 62. The wings scale with the peak luminosity of the line. This is consistent with electron scattering of the core of the line and indicates the same source for the wings and the narrow profile.}
\label{wings}
\end{minipage}
\end{figure}

In Fig.~\ref{wings}, we show continuum-subtracted and peak-normalised Balmer line profiles for day 62. We note that the \FeII\ lines have an effect on the H$\beta$ and H$\gamma$ profiles. DHG09 found the broad base of the Balmer line profiles of SN~1994W increase in strength when moving to shorter wavelengths. This effect is not as clear in the case of SN~2009kn. DHG09 also simulated the behaviour of the Paschen lines for SN~1994W-like objects. However, in our NIR spectra of SN~2009kn the Pa$\alpha$ line profile lands between the \textit{H}- and \textit{K}-band telluric line regions, and the signal-to-noise ratio for the rest of the Paschen lines is not high enough for a meaningful comparison.

\begin{figure}
\begin{minipage}{85mm}
\includegraphics[width=85mm, clip]
{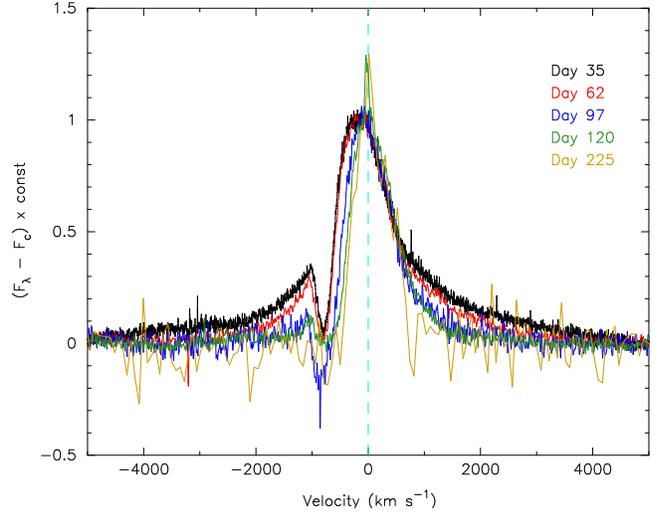}
\caption{The continuum-subtracted and peak-normalised H$\alpha$ profiles from UVES and X-Shooter overplotted. The disappearance of the electron scattering wings is consistent with a gradual thinning of the shell optical depth.}
\label{halpha}
\end{minipage}
\end{figure}

The broad base of the line profiles, shown most clearly by H$\alpha$, becomes weaker over time during the plateau phase and disappears after the light-curve drop. The same effect is observed also for the absorption component, whereas the narrow P Cygni emission component stays fairly constant throughout the whole observed line evolution. This indicates that the optical depth of the electron scattering is going down over time; this effect is clearly shown in Fig.~\ref{halpha}. 

\begin{figure*}
\includegraphics[width=130mm, clip]
{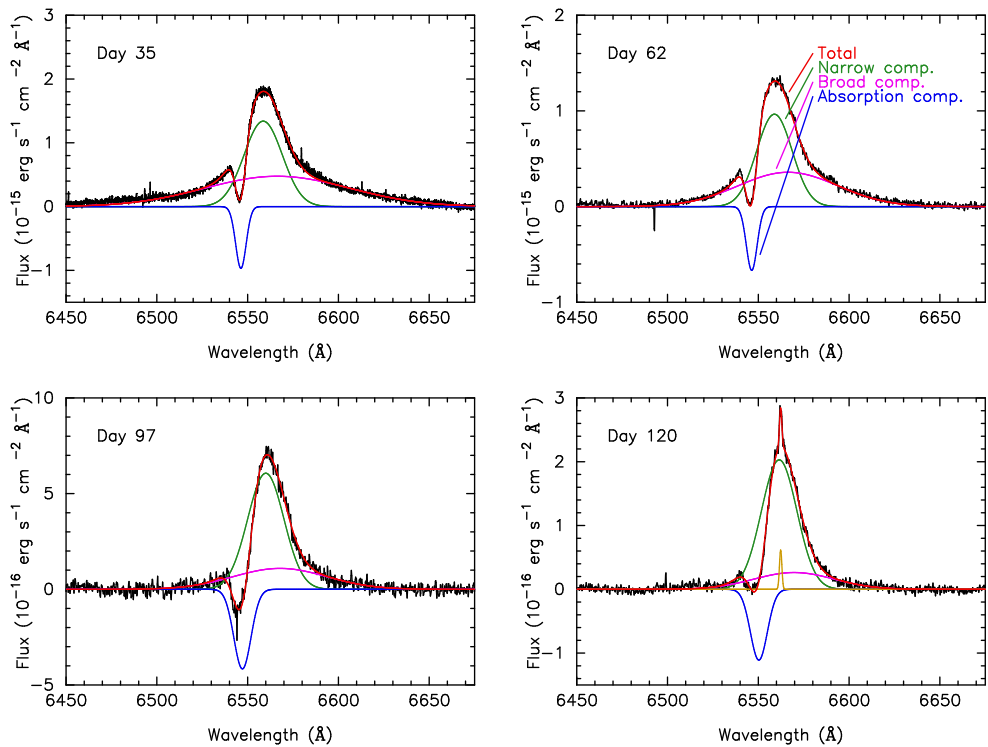}
\caption{Gaussian component fits for the continuum-subtracted H$\alpha$ profiles of SN~2009kn observed with VLT. The components are labelled in the day 62 spectrum, i.e. the narrow emission, the narrow absorption and the broad emission component. The narrow emission and absorption components form the narrow P Cygni line profile. An additional `very narrow' emission component is also seen in the day 120 spectrum.}
\label{gauss}
\end{figure*}

\begin{figure}
\begin{minipage}{85mm}
\includegraphics[width=85mm, clip]
{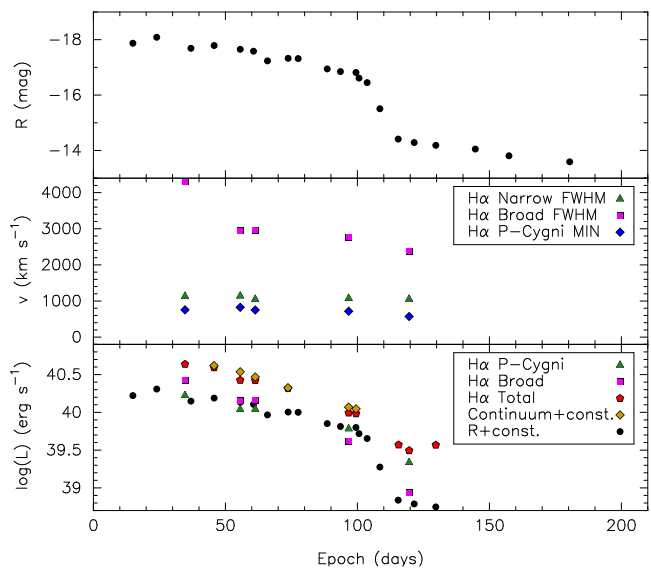}
\caption{Top: the \textit{R}-band light curve of SN~2009kn plotted as reference. Middle: instrumental resolution corrected velocity of different Gaussian components fitted for the H$\alpha$ line profile as a function of time. These include the FWHM of both the narrow and the broad Gaussian emission components and the position of the Gaussian absorption minimum based on the component fitting. Lower: the same as in the middle panel but for luminosity. These include the narrow P Cygni component, the broad emission component, the total H$\alpha$ line luminosity and a scaled continuum luminosity. See text and Fig.~\ref{gauss} for details.}
\label{comp}
\end{minipage}
\end{figure}

To investigate the evolution of different line components, we use a simple approach to analyse the line profiles. We fitted multiple Gaussian components to the higher resolution spectra of SN~2009kn, including the VLT spectra and day 56 NOT spectrum, using the external graphical {\sc spectool} task in {\sc iraf} (see Fig.~\ref{gauss}). For the time series of velocity and luminosity of different Gaussian components, see Fig.~\ref{comp}. Table~7 shows the measured velocities. We found a combination of a narrow P Cygni profile consisting of a Gaussian emission and absorption component and a broader Gaussian emission component to approximately describe the overall profile of H$\alpha$ in the plateau phase of SN~2009kn. These components can be seen in all the medium-resolution spectra through the whole observed evolution of SN~2009kn, with the narrow emission component having a roughly constant FWHM velocity of $\sim$1100~km~s$^{-1}$. All the velocity measurements are corrected for instrumental resolution. The velocity derived from the flux minimum of the narrow Gaussian absorption component decreases as a function of time from $\sim$800~km~s$^{-1}$ at day 35 to $\sim$600~km~s$^{-1}$ at day 120. Similarly, the FWHM velocity of the broad emission component decreases from $\sim$3500~km~s$^{-1}$ at day 35 to $\sim$2400~km~s$^{-1}$ at day 120. In the case of the day 120 tail phase X-Shooter spectrum, we found it necessary to add one additional extremely narrow emission component with a resolved FWHM velocity of $\sim$60~km~s$^{-1}$, associated with an \HII\ region as mentioned in Section~4.3. The luminosities of the narrow P Cygni and the broad emission components as well as the total H$\alpha$ luminosities were found to follow well the evolution of the \textit{R}-band luminosity in the plateau phase. However, in the tail phase only the luminosity of the broad emission component follows the \textit{R}-band light curve, whereas the total luminosity and narrow P Cygni component (which dominates the total luminosity) do not drop as steeply. 

\begin{table}
\begin{center}
\caption{Instrumental resolution corrected velocities derived from the best Gaussian component fits to the H$\alpha$ profile.}
\begin{tabular}{ccccc}
\hline
Epoch & Narrow FWHM & Broad FWHM & P Cygni min\\ 
(d) & (km~s$^{-1}$) & (km~s$^{-1}$) & (km~s$^{-1}$) \\ \hline
35 & 1133 & 4298 & 755 \\
56 & 1140 & 2949 & 825 \\
62 & 1050 & 2954 & 751 \\
97 & 1075 & 2770 & 717 \\
120 & 1051 & 2376 & 571 \\
\hline
\end{tabular}
\end{center}
\end{table}

\section{Discussion}

\subsection{Previous models for SN~1994W}

Here we discuss the two models presented for SN~1994W. CBC04 conducted both radiative transfer and hydrodynamical modeling, which they compared with the observed spectra and light curves of SN~1994W. They concluded that SN~1994W was surrounded by a dense $\sim$0.4~M$_{\odot}$ circumstellar (CS) envelope with an outer radius of $\sim$3~$\times$~10$^{15}$~cm, ejected in a violent outburst by the SN progenitor $\sim$1.5~yr before core collapsing and expanding with a velocity of $\sim$1000~km~s$^{-1}$. CBC04 identified three components in the observed line profiles during the plateau phase of SN~1994W. The first component is a narrow P Cygni line with $700$~km~s$^{-1}$ absorption minima arising from the expanding CS envelope. The second is a broad emission component with maximum blue velocity $\sim$4000~km~s$^{-1}$ originating from the CDS. The broad wings extending to $-5000$~km~s$^{-1}$ in the blue and $+7000$~km~s$^{-1}$ in the red make the third component, which is attributed to multiple Thomson scattering in the expanding CS envelope. Due to the high optical depth of the CDS, formed at the interface of the SN ejecta and CS envelope, only the photosphere can be observed, thus hiding any broad P Cygni emission lines arising in the hydrogen-rich ejecta. Therefore, CBC04 interpret the $\sim$4000~km~s$^{-1}$ velocity of the broad emission line component to correspond to the expansion velocity of the SN ejecta. The interaction of the SN ejecta with dense CSM gives rise to a higher SN luminosity than observed for typical Type IIP SNe, and the high-density medium slows down the SN ejecta. A similar model of a dense CS envelope ejected by the SN progenitor in an outburst a few years before core collapse was also suggested for SN~1995G by \citet{chugai03}.

Our late tail phase observations of SN~2009kn make it difficult to understand how the narrow-line component could be associated with the CSM envelope and the broad component with the SN ejecta as presented by CBC04. With our extensive data set for SN~2009kn, we can see that coincident with the light-curve drop the broad component begins to fade rapidly, while the narrow component persists. Based on the model of CBC04, which associates the broad component with the ejecta, we would expect to continue observing this component also after the light-curve drop. Neither do the observations support the idea that the ejecta swept up and accelerated the entire CSM shell by the end of the plateau unless this took place at a much earlier epoch, before our first spectroscopic observations. Moreover, we observe no dramatic transformation of the SN~2009kn spectrum from the plateau to the tail (on day 120). This suggests that radiation transfer from the shell interior dominates. We suggest that the drop in the light curve marks not the ejecta reaching the outer edge of the entire CSM envelope, but instead the end of the cooling phase just as seen in Type IIP SNe. 

DHG09 conducted a spectroscopic analysis of SN~1994W based on radiative transfer modelling, without performing any hydrodynamical modeling of the light curves. They used the model to independently produce synthetic counterparts for six epochs of observed spectra of SN~1994W. DHG09 note that their model fits poorly to the early spectra of the brightening phase of days 21 and 31. However, the following four epochs during the plateau phase reproduce the observed spectra well. DHG09 concluded that the entire line profile can be explained by a single optically thick emitting region, where the broad component results from internal multiple electron scattering in the photosphere. For comparison, in the model of CBC04, a CDS forms as a result of shock interaction between the ejecta and the CS envelope, and it behaves subsequently as a photosphere and remains optically thick throughout the plateau phase. DHG09 explain the end of the plateau phase and the drop in the light curve as the point when the photosphere becomes optically thin after receding into the fully recombined central parts of the ejecta, whereas in the model of CBC04 the drop happens when the photosphere breaks through the more slowly expanding CS envelope. DHG09 suggests reverse-shock deceleration as the reason for the lack of broad lines observed during the nebular phase. However, based on the lack of broad nebular lines and the low mass of $^{56}$Ni in SN~1994W, DHG09 also proposed that it is not necessary for the inner shell to originate from core collapse. Instead, it might have been ejected in another outburst by the progenitor star which, due to its higher velocity, catches up with the expanding outer shell. The resulting interaction would give the observed display. They propose a model where the two shells have similar masses, so that a strong enough deceleration of the inner shell could be produced.

\subsection{The nature of SN~2009kn}

As the bolometric tail phase luminosity of SN~2009kn is consistent with the radioactive decay tail, this supports an SN origin for SN~2009kn. This is further supported by observations of other SN~2009kn-like events. For SN~1999el, we estimate a pseudo-bolometric decline rate of $\gamma=0.92\pm0.06$~mag (100 d)$^{-1}$ between epochs 113 and 228 based on the \textit{UBVRIJHK} light curves of \citet{dicarlo02} (assuming $A_V=1.18$~mag; see Section~3.3). For SN~1999eb, \citet{li02} found a late time \textit{I}-band decline rate of $\gamma_{I}=0.86\pm0.01$~mag (100 d)$^{-1}$ using data on epochs 332, 387 and 659.

With a suitable selection of parameters, an interaction between the shells ejected in two massive and consecutive outbursts could still create a tail phase decline that mimics radioactive decay. This possibility was addressed by DHG09 in the case of SN~1994W, for which the available data (in \textit{R}) showed faster decline than expected from radioactivity. This might also explain why such different light curves, resembling both Type IIP and IIL SNe, are shown by events with similar spectra \citep[see also][]{dessart10}. If the tail phase decline is due not to radioactive decay but to interaction between shells produced by outbursts of the progenitor, this would also mean that a similar coincidence is seen in both SN~1999eb and SN~1999el. The presence of strong forbidden nebular spectral lines, particularly the [\OI] $\lambda\lambda$~6300.30,~6363.78 blend and \CaII] $\lambda\lambda$~7291.47,~7324.89, would support a core-collapse origin for SN~2009kn, but we only see a hint of the \CaII] blend and [\OI] is absent in the late-time spectra. If SN~2009kn was indeed an SN, we associate the faintness of the forbidden lines with an effect of high density causing these lines to be collisionally de-excited \citep[see e.g.][]{filippenko89, chugai94, benetti98, fransson02}.

\begin{figure}
\begin{minipage}{85mm}
\includegraphics[width=85mm, clip]
{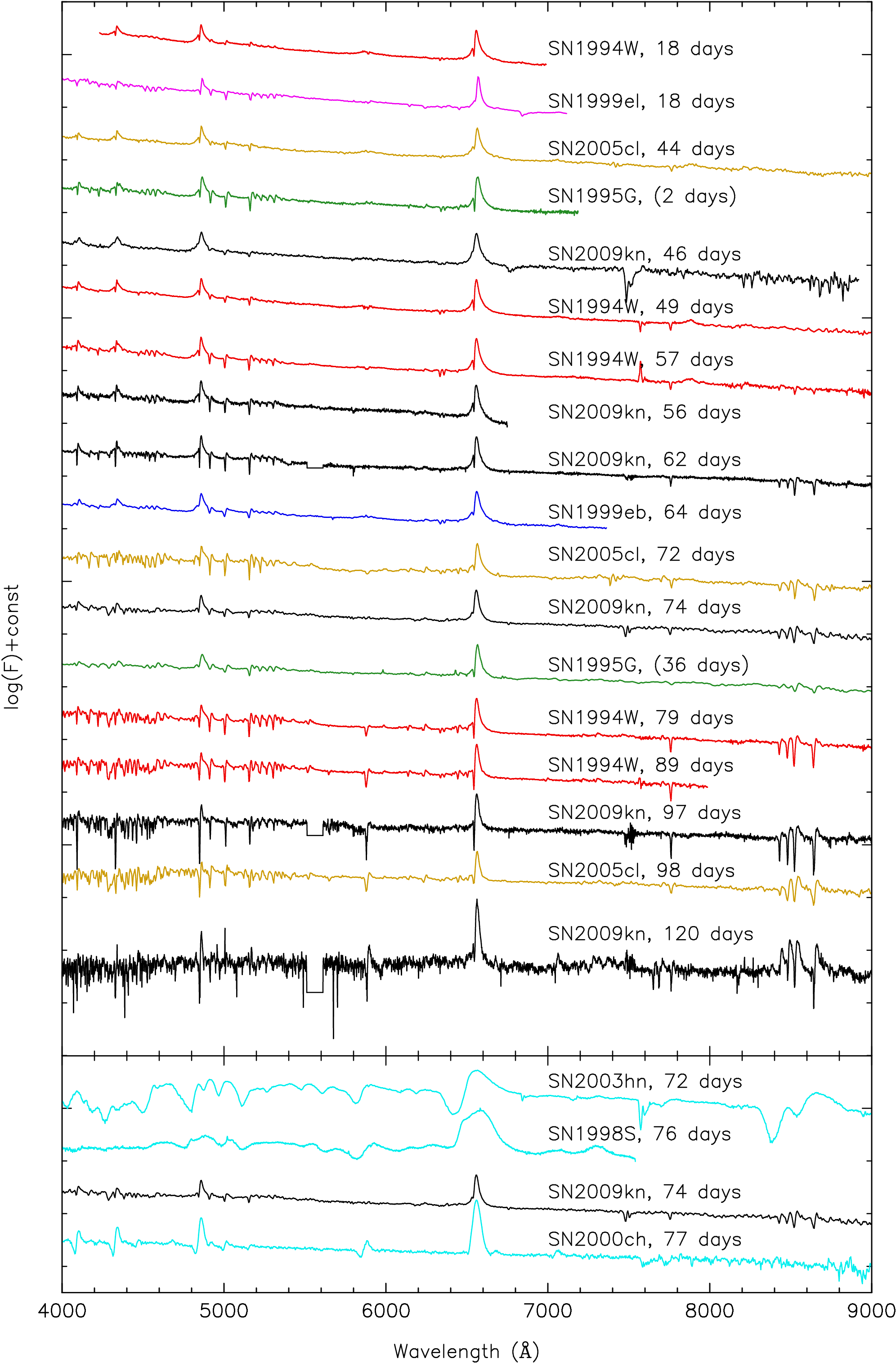}
\caption{Top: time series comparison of SN~2009kn spectra with other Type IIn SNe with very narrow Balmer lines $-$ SN~1994W (CBC04), SN~1995G, SN~1999eb, SN~1999el \citep{pastorello02} and SN~2005cl$^{10}$ \citep{kiewe12}. The epochs of the SN~1995G spectra are after discovery and are shown in parentheses. Bottom: the day 74 spectrum of SN~2009kn compared to the Type IIP SN~2003hn \citep{krisciunas09}, the Type IIn SN~1998S and SN~2000ch, an SN impostor in NGC 3234 after its 2008 outburst \citep{pastorello10}. The spectra have all been dereddened with wavelengths corrected to the host galaxy rest frame and vertically shifted for clarity.}
\label{time}
\end{minipage}
\end{figure}

When comparing the spectral time series of SN~2009kn-like SNe, we find a striking similarity in the spectral evolution within this family of Type IIn SNe (see Fig.~\ref{time}). All the SN~2009kn-like events have optical absolute peak magnitudes of $-18$ to $-19$~mag, which are relatively bright for being a Type II SN. CSM interaction likely explains these bright peak luminosities. A few other comparison spectra are also plotted around epoch 70$-$80~d showing the clear difference to the Type IIn SN~1998S and the Type IIP SN~2003hn. We also compare SN~2009kn with the LBV outburst SN~2000ch, an SN impostor. The spectrum of SN~2000ch after its 2008 outburst has similar narrow-line profiles but it is much less luminous than SN~2009kn \citep{pastorello10}. 

As shown in Section~4.5 for the H$\alpha$ line components, the luminosity of the broad-line component seems to closely follow the luminosity of the narrow-line component during the plateau phase. This, together with the symmetry of the broad component, supports the conclusion that both the broad-line and the narrow-line components arise from the same physical region, consistent with the model of DHG09 where the narrow component is associated with the photosphere at $\tau\sim1$ and the broad component is due to internal electron scattering. In other words, it would require careful fine tuning of the parameters to have the narrow component arise from the CSM and the broad component from the ejecta as suggested by CBC04 for SN~1994W if their luminosities are to scale together over time. Furthermore, the velocity width of the narrow component stays fairly constant during the plateau phase, while the width of the broad component decreases. This is consistent with the temperature dropping, reducing the strength of the electron scattering. For these reasons, we favor the model of DHG09 to explain the broad base of the spectral lines with electron scattering. 

\addtocounter{footnote}{1}
\footnotetext{Data from the Weizmann interactive supernova data repository \citep{yaron12}} 

As noted above, no broad spectral lines are detected in the tail phase of SN~2009kn that would indicate the higher velocities expected for typical optically thin SN ejecta. Momentum conservation argues for a CSM-ejecta mass ratio $\oldgeq$1 if the CSM is to have a significant effect on the ejecta velocity. If we assume that SN~2009kn is a genuine SN, we suggest the following scenario. The supernova ejecta collide with a massive CS envelope, forming a swept-up shell with a velocity of $v_s\sim800$~km~s$^{-1}$. The internal energy released in the collision is converted into thermal energy, which is essentially trapped inside the optically thick swept-up shell. In many respects, the cooling of this shell is similar to what occurs during the plateau phase in Type IIP SNe. One possible explanation for the difference in this case is that, due to the large radius of the CS envelope, the trapped radiation at any particular epoch is hotter than in the corresponding Type IIP case. The recombination wave in SN~2009kn might not have formed until the very end of the plateau. The drop in the end of the plateau phase is analogous to the drop shown by genuine Type IIP SNe. The associated disappearance of the photosphere ends internal electron scattering in it and the broad H$\alpha$ component disappears. 

Subluminous, SN~2005cs-like (see Section~3.2) Type IIP SNe are particularly interesting in comparison to SN~2009kn because of their low $^{56}$Ni masses and the narrow-line profiles indicating velocities similar to SN~2009kn of the order of 1000~km~s$^{-1}$ at day $\sim$100. However, as shown by \citet{pastorello04, pastorello06, pastorello09} these low-luminosity Type IIP SNe still have higher velocities, of $\sim$4000~km~s$^{-1}$ on early epochs in contrast to SN~2009kn. In addition, the $^{56}$Ni masses derived for these SNe are lower than our estimate for SN~2009kn. For example \citet{pastorello09} gave a $^{56}$Ni mass estimate of $0.003M_{\odot}$ for SN~2005cs, which is an order of magnitude below the inferred $^{56}$Ni mass for SN~2009kn. Based on direct pre-explosion progenitor observations, \citet{maund05}, \citet{li06} and \citet{eldridge07} found SN~2005cs to have originated from a star with an initial mass in the range $\sim$6$-$10~M$_{\odot}$. It seems unlikely that a red supergiant star of such a low mass would have experienced a massive outburst producing the CSM which SN~2009kn is interacting with; instead, SN 2009kn might have been a so-called electron-capture SN originating from a super-asymptotic giant branch (super-AGB) star close to the core-collapse limit $\sim$8M$_{\odot}$. Super-AGB stars have been excluded as the progenitor of several subluminous Type IIP SNe, such as SN~2005cs (Eldridge et al. 2007), SN~2008bk \citep{mattila08, vandyk12}, and SN~2009md \citep{fraser11}. However, super-AGB stars have been suggested as the progenitors of fainter ($-12 \lesssim M_{R,max} \lesssim -15$) events such as SN~2008S, SN~2008ha and the 2008 optical transient in NGC 330 \citep[see][]{prieto08, botticella09, pumo09, thompson09, valenti09}. Similarly, a super-AGB progenitor has also been suggested for the bright ($M_{R,max}\approx -18$) Type IIP SN~2007od with a low derived $^{56}$Ni mass \citep{inserra11}. The spectra of SN~2009kn and SN~1994W are nevertheless quite unlike these events. 

In the electron-capture SN model, core collapse is triggered by electron capture of $^{24}$Mg and $^{20}$Ne, which reduces the electron pressure and leads to a collapse of the oxygen-neon core, instead of an iron core \citep[see e.g.][]{miyaji80, nomoto84}. The hydrogen envelope is thought to be only loosely tied to the super-AGB star and a significant amount of mass could be ejected in an outburst due to a thermal instability \citep{weaver79}. If such major mass loss episode takes place shortly before the core collapse, it could lead to an SN showing a strong ejecta-CSM interaction. The simulations of \citet{kitaura06} indicate a very low ($<0.015M_{\odot}$) mass of $^{56}$Ni for electron capture SNe, which is inferred in SN~2009kn. \citet{weaver79} suggest a maximum velocity of the ejected hydrogen envelope of around $\sim$100~km~s$^{-1}$. As shown in Section~4.3, the good resolution of our VLT spectra of SN~2009kn would have enabled us to detect and resolve any line components as narrow as this. However, we see no narrower component than the dominating P Cygni lines in our UVES and X-Shooter spectra. The original velocity of the CSM cannot have been much different from the observed 800~km~s$^{-1}$ P Cygni velocity unless the CSM component is an unseen rarefied wind. As already suggested in the preliminary analysis of \citet{pastorello11}, the observed properties of SN~2009kn agree overall well with a model of an O/Ne/Mg core collapsing, preceded by a CSM envelope forming in Ne flash driven outbursts (see also the discussion by CBC04). We do point out that the $\sim$1000~km~s$^{-1}$ velocity we see is also consistent with a very massive LBV-like outburst \citep[e.g.][]{pastorello10}. However, if SN~2009kn is considered as a core-collapse event, it is problematic to associate an LBV as the progenitor since they should not explode as SNe according to the current stellar evolution models \citep[e.g.][]{langer94}. 

\section{Conclusions}

We have shown SN~2009kn to be a twin of SN~1994W with only a few differences. The two events have similar light curves and colour evolution during a plateau phase. SN~1994W was roughly 1~mag brighter and had a few days longer plateau than SN~2009kn. The plateau phase was followed by a deep light-curve drop, slightly deeper in the case of SN~1994W, followed by a tail phase. The slope of the tail phase is the largest observed difference between the two SNe. SN~2009kn follows fairly well the radioactive decay tail of $^{56}$Co, which supports an SN nature for SN~2009kn. We infer an upper limit for the $^{56}$Ni mass of $0.023$~M$_{\odot}$, much higher than the estimate for SN~1994W. However, given the rapid decline of the light curve of SN~1994W at the tail phase, the possibility remains that an energy source other than radioactive decay (e.g. CS interaction) could dominate in both supernovae at the tail phase. Moreover, we cannot rule out that the energy sources for the tail luminosity in these SNe are different, $^{56}$Co dominating in SN~2009kn and some other energy source dominating in SN~1994W. Therefore, the two objects, SN~2009kn and SN~1994W, could even be different kinds of transient events originating from different types of progenitors, as the CSM interaction in both the objects explains the observed similarities before the post-plateau light-curve drop. For SN~2009kn, we detect late-time NIR emission which most likely arises from newly formed dust. Unfortunately, no NIR data are available for SN~1994W for further comparison. 

Our systematically obtained optical and NIR multiband photometry together with the VLT/X-Shooter spectra as late as the tail phase makes our data set the most comprehensive data set available for future detailed modelling of SN~1994W-like events. The comparison between the spectral evolution of SN~2009kn and SN~1994W also reveals striking similarity in both evolution of the line luminosities and the details of their profiles. The \CaII] and [\OI] lines which are typically strong in the nebular spectra of normal Type II SNe are faint or absent in the tail phase spectra of SN~2009kn. We attribute this to high density in the ejecta. Our spectra excludes the existence of very narrow P Cygni lines that could be associated with slow progenitor wind. Furthermore, we emphasize the similarity of the spectral evolution of other SN~2009kn- and SN~1994W-like Type IIn SNe, even though they do not necessarily show the similar Type IIP-like light curve. We have shown that the spectral line components of SN~2009kn are in a good agreement with the model of DHG09, where the broad wings in particular are caused by internal electron scattering and the whole line profile arises from the same physical region. 

\section*{Acknowledgments}

We thank the anonymous referee for very useful comments and suggestions. We thank Luc Dessart for very helpful comments and discussions. We thank Stefan Taubenberger for obtaining the Calar Alto observations and Irene Agnoletto for carrying out some of the NTT observations.

EK acknowledges financial support from the Finnish Academy of Science and Letters (Vilho, Yrj\"{o} and Kalle V\"{a}is\"{a}l\"{a} Foundation). SM and EK acknowledge financial support from the Academy of Finland (project: 8120503). AP, SB, MT and SV are partially supported by the PRIN-INAF 2009 with the project `Supernovae Variety and Nucleosynthesis Yields'. GL is supported by a grant from the Carlsberg foundation. DARK is supported by the DNRF.

This paper is based on observations made with the Nordic Optical Telescope, operated on the island of La Palma jointly by Denmark, Finland, Iceland, Norway and Sweden, in the Spanish Observatorio del Roque de los Muchachos of the Instituto de Astrofisica de Canarias

The Liverpool Telescope is operated on the island of La Palma by Liverpool John Moores University in the Spanish Observatorio del Roque de los Muchachos of the Instituto de Astrofisica de Canarias with financial support from the UK Science and Technology Facilities Council. 

This work is based on observations made with ESO Telescopes at the La Silla and Paranal Observatories under programme IDs 184.D-1140, 084.D-0265 and 084.D-0463.

This paper is based on observations collected at the Centro Astron\'{o}mico Hispano Alem\'{a}n (CAHA) at Calar Alto, operated jointly by the Max-Planck Institut f\"ur Astronomie and the Instituto de Astrof\'{i}sica de Andaluc\'{i}a (CSIC).

The LBT is an international collaboration among institutions in the US, Italy and Germany. LBT Corporation partners are The University of Arizona on behalf of the Arizona university system, Istituto Nazionale di Astrofisica (Italy), LBT Beteiligungsgesellschaft (Germany), representing the Max-Planck Society, the Astrophysical Institute Potsdam and Heidelberg University, The Ohio State University and The Research Corporation on behalf of The University of Notre Dame, University of Minnesota and University of Virginia.

This work has been greatly facilitated and expedited by the European supernova collaboration involved in ESO-NTT large program 184.D-1140 led by Stefano Benetti.

We are grateful to the amateur astronomers from the Puckett Observatory Supernova Search for providing us their original observations.

This research has made use of the NASA/IPAC Extragalactic Database (NED) which is operated by the Jet Propulsion Laboratory, California Institute of Technology, under contract with the National Aeronautics and Space Administration.

We acknowledge the usage of the HyperLeda data base (http://leda.univ-lyon1.fr).

We have made use of the Weizmann interactive supernova data repository (www.weizmann.ac.il/astrophysics/wiserep).

\appendix

\section{Tables}

Identified lines in the spectra of SN~2009kn. 

\newpage

\begin{table*}
\begin{center}
\caption{Identified lines in the optical spectra of SN~2009kn. We use the same symbols as in CBC04, i.e. 'e' for emission line and 'p' for P Cygni line.}
\begin{tabular}{lccccccc}
\hline
 & \multicolumn{7}{c}{Epoch (d)} \\ 
Line (\AA)& 46 & 56 & 62 & 74 & 97 & 120 & 130\\ \hline

$[$\OII$]$ 3726.03 + $[$\OII$]$ 3728.82 & - & - & - & - & - & e & \\
H11 3770.63 & p & - & p & p & - & - & \\
H10 3797.90 & - & p & p & p & - & - & \\
H9 3835.38 & - & - & p & p & p & - & \\
3880.? & - & - & p & - & - & - & \\
H8 3889.05 & p & p & p & p & p & - & \\
\TiII\ 3900.55 & - & - & p & - & - & - & \\
\TiII\ 3913.47 & - & - & p & - & - & - & \\
\CaII\ 3933.66 & - & p & p & p & p & - & \\
\CaII\ 3968.47 + H$\epsilon$ 3970.07 & - & p & p & p & p & - & \\
H$\delta$ 4101.74 & e & p & p & p & p & - & \\
\FeII\ 4178.85 & - & p & - & p & - & - & \\
\FeII\ 4233.12 & e & e & p & p & p & - & \\
\ScII\ 4246.82 & - & - & - & - & p & - & \\
H$\gamma$ 4340.47 & e & p & p & p & p & p & \\
\FeII\ 4351.76 & - & e & e & - & - & - & \\
\TiII\ 4395.03 & - & - & p & p & p & - & \\
\FeII\ 4416.83 & - & - & - & p & p & - & \\
\TiII\ 4443.79 & - & - & - & p & p & - & \\
\TiII\ 4468.51 & - & - & - & p & p & - & \\
\HeI\ 4471.68 & - & - & - & - & - & e & \\
\TiII\ 4501.27 & - & - & - & - & p & - & \\
\FeII\ 4520.21 + \FeII\ 4522.63 & - & p & p & - & - & - & \\
\TiII\ 4533.97 & - & - & - & - & p & - & \\
\TiII\ 4549.62 & - & - & - & - & p & - & \\
\FeII\ 4555.88 & - & p & p & - & - & - & \\
\FeII\ 4582.83 + \FeII\ 4583.82 & - & p & p & p & - & - & \\
\FeII\ 4629.34 & - & p & - & e & - & - & \\
H$\beta$ 4861.33 & e & p & p & p & p & p & \\
\FeII\ 4923.94 & p & p & p & p & p & - & \\
$[$\OIII$]$ 4958.91 & - & - & - & - & - & e & \\
$[$\OIII$]$ 5006.84 & - & - & - & - & - & e & \\
\FeII\ 5018.44 & p & p & p & p & p & e & \\
\FeII\ 5169.05 & p & p & p & p & p & p & \\
\FeII\ 5234.63 + \ScII\ 5239.81 & - & p & p & p & p & - & \\
\FeII\ 5275.98 + \FeII\ 5284.12 & - & p & p & e & p & - & \\
\FeII\ 5316.78 & - & p & p & p & p & - & \\
\NaI\ 5889.95 + \NaI\ 5895.92 & - & - & - & - & p & p & \\
H$\alpha$ 6562.85 & e & p & p & p & p & p & e \\
\HeI\ 6678.15 & - & - & - & - & - & e & e \\
6778.? & - & - & - & - & - & p & - \\
\HeI\ 7065.19 & - & & - & - & - & e & e \\
\CaII$]$ 7291.47 & - & & - & - & - & e & e \\
\CaII$]$ 7323.89 & - & & - & - & - & e & e \\
\KI\ 7667.01 & - & & - & - & - & p & p \\
\KI\ 7701.08 & - & & - & - & - & p & p \\
\OI\ 7775.39 & - & & p & p & p & p & p \\
\OI\ 8446.76 & - & & p & p & p & e & e \\
\CaII\ 8498.02 & - & & p & p & p & p & e \\
\CaII\ 8542.09 & - & & p & p & p & p & e \\
\CaII\ 8662.14 & - & & p & p & p & p & p \\
Pa11 8862.79 & - & & - & - & p & - & - \\
Pa9 9229.01 & - & & - & - & p & p & - \\
Pa$\epsilon$ 9545.97 & - & & - & - & p & - & - \\
\hline
\end{tabular}
\end{center}
\end{table*}

\begin{table}
\begin{center}
\caption{Identified lines in the NIR spectra of SN~2009kn. We use the same symbols as in CBC04, i.e. 'e' for emission line and 'p' for P Cygni line.}
\begin{tabular}{lcccc}
\hline
 & \multicolumn{4}{c}{Epoch (d)} \\
Line (\AA) & 62 & 76 & 97 & 120$^{a}$ \\ \hline

Pa$\delta$ 10049.40 & - & e & - & e \\
\HeI\ 10830.25 + \HeI\ 10830.34  & - & - & p & p \\
Pa$\gamma$ 10938.10 & p & e & p & p \\
\OI\ 11286.91 & - & - & - & e \\
Pa$\beta$ 12818.10 & p & e & p & e \\
\MgI\ 15025.00 + \MgI\ 15040.25  & - & - & - & e \\
\hline
\end{tabular}
\end{center}
$^{a}$ On epoch 123, our automatic method described in Section~4.3 recognizes additional faint features roughly at 10236, 12707 and 15321\AA\ as lines. However, we do not identify them and associate their detection as lines arising most likely from problems in the used method. 
\end{table}


\begin{thebibliography}{}

\bibitem[Aretxaga et al.(1999)]{aretxaga99} Aretxaga, I., Benetti S., Terlevich R.~J., Fabian A.~C., Cappellaro E., Turatto M., della Valle M.,\ 1999, \mnras, 309, 343 
\bibitem[Benetti et al.(1998)]{benetti98} Benetti S., Cappellaro E., Danziger I.~J., Turatto M., Patat F., della Valle M.,\ 1998, \mnras, 294, 448 
\bibitem[Blondin \& Tonry(2007)]{blondin07} Blondin S., Tonry J.~L.,\ 2007, \apj, 666, 1024 
\bibitem[Botticella et al.(2009)]{botticella09} Botticella M.~T. et al.,\ 2009, \mnras, 398, 1041 
\bibitem[Bouchet et al.(1991)]{bouchet91} Bouchet P., Phillips M.~M., Suntzeff N.~B., Gouiffes C., Hanuschik R.~W., Wooden D.~H.,\ 1991, \aap, 245, 490 
\bibitem[Buzzoni et al.(1984)]{buzzoni84} Buzzoni B. et al.,\ 1984, The Messenger, 38, 9 
\bibitem[Cardelli, Clayton \&  Mathis(1989)]{cardelli89} Cardelli J.~A., Clayton G.~C., Mathis J.~S.,\ 1989, \apj, 345, 245 
\bibitem[Chamaraux et al.(1999)]{chamaraux99} Chamaraux P., Masnou J.-L., Kaz{\'e}s I., Sait{\= o} M., Takata T., Yamada T.,\ 1999, \mnras, 307, 236 
\bibitem[Chandra \& Soderberg(2009)]{chandra09} Chandra P., Soderberg A.,\ 2009, The Astron. Telegram, 2335, 1
\bibitem[Chugai \& Danziger(1994)]{chugai94} Chugai N.~N., Danziger I.~J.,\ 1994, \mnras, 268, 173 
\bibitem[Chugai \& Danziger(2003)]{chugai03} Chugai N.~N., Danziger I.~J.,\ 2003, Astron. Lett., 29, 649 
\bibitem[Chugai et al.(2004)]{chugai04} Chugai N.~N. et al.,\ 2004, \mnras, 352, 1213 (CBC04)
\bibitem[Dekker et al.(2000)]{dekker00} Dekker H., D'Odorico S., Kaufer A., Delabre B., Kotzlowski H.,\ 2000, \procspie, 4008, 534 
\bibitem[Dessart et al.(2009)]{dessart09} Dessart L., Hillier D.~J., Gezari S., Basa S., Matheson T.,\ 2009, \mnras, 394, 21 (DHG09)
\bibitem[Dessart, Livne \& Waldman(2010)]{dessart10} Dessart L., Livne E., Waldman R.,\ 2010, \mnras, 405, 2113 
\bibitem[Di Carlo et al.(2002)]{dicarlo02} Di Carlo E. et al.,\ 2002, \apj, 573, 144 
\bibitem[Djupvik \& Andersen(2010)]{djupvik10} Djupvik A.~A., Andersen J.,\ 2010, in Diego J.~M., Goicoecha L.~J., Gonz\'alez-Serrano J.~I., Gorgas J., eds, Astrophys. Space Sci. Proc., Highlights of Spanish Astrophysics V. Springer-Verlag, Berlin, p.~211 
\bibitem[Eldridge, Mattila \& Smartt(2007)]{eldridge07} Eldridge J.~J., Mattila S., Smartt S.~J.,\ 2007, \mnras, 376, L52 
\bibitem[Elmhamdi, Chugai \& Danziger(2003)]{elmhamdi03} Elmhamdi A., Chugai N.~N., Danziger I.~J.,\ 2003, \aap, 404, 1077 
\bibitem[Fassia et al.(2000)]{fassia00} Fassia A. et al.,\ 2000, \mnras, 318, 1093 
\bibitem[Fassia et al.(2001)]{fassia01} Fassia A. et al.,\ 2001, \mnras, 325, 907 
\bibitem[Filippenko(1989)]{filippenko89} Filippenko A.~V.,\ 1989, \aj, 97, 726 
\bibitem[Filippenko(1997)]{filippenko97} Filippenko A.~V.,\ 1997, \araa, 35, 309
\bibitem[Fransson et al.(2002)]{fransson02} Fransson C. et al.,\ 2002, \apj, 572, 350 
\bibitem[Fraser et al.(2011)]{fraser11} Fraser M. et al.,\ 2011, \mnras, 417, 1417 
\bibitem[Gagliano et al.(2009)]{gagliano09} Gagliano R., Newton J., Puckett T., Orff T.,\ 2009, Cent. Bureau Electron. Telegrams, 1997, 1 
\bibitem[Goldoni et al.(2006)]{goldoni06} Goldoni P., Royer F., Fran{\c c}ois P., Horrobin M., Blanc G., Vernet J., Modigliani A., Larsen J.,\ 2006, \procspie, 6269, 62692K
\bibitem[Inserra et al.(2011)]{inserra11} Inserra C. et al.,\ 2011, \mnras, 417, 261 
\bibitem[Jerkstrand, Fransson \& Kozma(2011)]{jerkstrand11} Jerkstrand A., Fransson C., Kozma C.,\ 2011, \aap, 530, A45 
\bibitem[Kiewe et al.(2012)]{kiewe12} Kiewe M. et al.,\ 2012, \apj, 744, 10
\bibitem[Kitaura, Janka \& Hillebrandt(2006)]{kitaura06} Kitaura F.~S., Janka H.-T., Hillebrandt W.,\ 2006, \aap, 450, 345 
\bibitem[Kotak et al.(2005)]{kotak05} Kotak R., Meikle P., van Dyk S.~D., H{\"o}flich P.~A., Mattila S.,\ 2005, \apjl, 628, L123 
\bibitem[Kotak et al.(2009)]{kotak09} Kotak R. et al.,\ 2009, \apj, 704, 306 
\bibitem[Krisciunas et al.(2009)]{krisciunas09} Krisciunas K. et al.,\ 2009, \aj, 137, 34 
\bibitem[Landolt(1992)]{landolt92} Landolt A.~U.,\ 1992, \aj, 104, 340 
\bibitem[Langer et al.(1994)]{langer94} Langer N., Hamann W.-R., Lennon M., Najarro F., Pauldrach A.~W.~A., Puls J.,\ 1994, \aap, 290, 819 
\bibitem[Li et al.(2002)]{li02} Li W., Filippenko A.~V., Van Dyk S.~D., Hu J., Qiu Y., Modjaz M., Leonard D.~C.,\ 2002, \pasp, 114, 403 
\bibitem[Li et al.(2006)]{li06} Li W., Van Dyk S.~D., Filippenko A.~V., Cuillandre J.-C., Jha S., Bloom J.~S., Riess A.~G., Livio M.,\ 2006, \apj, 641, 1060 
\bibitem[Li et al.(2011)]{li11} Li W. et al.,\ 2011, \mnras, 412, 1441 
\bibitem[Liu et al.(2000)]{liu00} Liu Q.-Z., Hu J.-Y., Hang H.-R., Qiu Y.-L., Zhu Z.-X., Qiao Q.-Y.,\ 2000, \aaps, 144, 219 
\bibitem[Mandel et al.(2000)]{mandel00} Mandel H. et al.,\ 2000, \procspie, 4008, 767 
\bibitem[Mattila et al.(2008)]{mattila08} Mattila S., Smartt S.~J., Eldridge J.~J., Maund J.~R., Crockett R.~M., Danziger I.~J.,\ 2008, \apjl, 688, L91 
\bibitem[Maund, Smartt \& Danziger(2005)]{maund05} Maund J.~R., Smartt S.~J., Danziger I.~J.,\ 2005, \mnras, 364, L33 
\bibitem[Mazzali \& Lucy(1993)]{mazzali93} Mazzali P.~A., Lucy L.~B.,\ 1993, \aap, 279, 447 
\bibitem[Meikle et al.(2006)]{meikle06} Meikle W.~P.~S. et al.,\ 2006, \apj, 649, 332 
\bibitem[Meikle et al.(2011)]{meikle11} Meikle W.~P.~S. et al.,\ 2011, \apj, 732, 109 
\bibitem[Meisenheimer (1998)]{meisenheimer98} Meisenheimer K.,\ 1998, User Guide to the CAFOS 2.2
\bibitem[Miyaji et al.(1980)]{miyaji80} Miyaji S., Nomoto K., Yokoi K., Sugimoto D.,\ 1980, \pasj, 32, 303 
\bibitem[Moorwood, Cuby \& Lidman(1998)]{moorwood98} Moorwood A., Cuby J.-G., Lidman C.,\ 1998, The Messenger, 91, 9 
\bibitem[Mould et al.(2000)]{mould00} Mould J.~R. et al.,\ 2000, \apj, 529, 786 
\bibitem[Nomoto(1984)]{nomoto84} Nomoto K.,\ 1984, \apj, 277, 791 
\bibitem[Osterbrock \& Ferland(2006)]{osterbrock06} Osterbrock D.~E., Ferland G.~J.,\ 2006, Astrophysics of gaseous nebulae and active galactic nuclei, 2nd~edn. University Science Books, Sausalito, CA 
\bibitem[Pastorello et al.(2002)]{pastorello02} Pastorello A. et al.,\ 2002, \mnras, 333, 27 
\bibitem[Pastorello et al.(2004)]{pastorello04} Pastorello A. et al.,\ 2004, \mnras, 347, 74 
\bibitem[Pastorello et al.(2006)]{pastorello06} Pastorello A. et al.,\ 2006, \mnras, 370, 1752 
\bibitem[Pastorello et al.(2009)]{pastorello09} Pastorello A. et al.,\ 2009, \mnras, 394, 2266 
\bibitem[Pastorello et al.(2010)]{pastorello10} Pastorello A. et al.,\ 2010, \mnras, 408, 181 
\bibitem[Pastorello et al.(2011)]{pastorello11} Pastorello A., Benetti S., Bufano F., Kankare E., Mattila S., Turatto M., Cupani G.,\ 2011, Astron. Nachr., 332, 266 
\bibitem[Paturel et al.(2003)]{paturel03} Paturel G., Petit C., Prugniel P., Theureau G., Rousseau J., Brouty M., Dubois P., Cambr{\'e}sy L.,\ 2003, \aap, 412, 45 
\bibitem[Pickles(1998)]{pickles98} Pickles A.~J.,\ 1998, \pasp, 110, 863 
\bibitem[Pignata et al.(2011)]{pignata11} Pignata G. et al.,\ 2011, Cent. Bureau Electron. Telegrams, 2623, 1 
\bibitem[Poznanski et al.(2011)]{poznanski11} Poznanski D., Ganeshalingam M., Silverman J.~M., Filippenko A.~V.,\ 2011, \mnras, 415, L81 
\bibitem[Pozzo et al.(2004)]{pozzo04} Pozzo M., Meikle W.~P.~S., Fassia A., Geballe T., Lundqvist P., Chugai N.~N., Sollerman J.,\ 2004, \mnras, 352, 457 
\bibitem[Pozzo et al.(2006)]{pozzo06} Pozzo M. et al.,\ 2006, \mnras, 368, 1169 
\bibitem[Pradhan et al.(2006)]{pradhan06} Pradhan A.~K., Montenegro M., Nahar S.~N., Eissner W.,\ 2006, \mnras, 366, L6 
\bibitem[Prieto et al.(2008)]{prieto08} Prieto J.~L. et al.,\ 2008, \apjl, 681, L9 
\bibitem[Pumo et al.(2009)]{pumo09} Pumo M.~L. et al.,\ 2009, \apjl, 705, L138 
\bibitem[Ralchenko et al.(2008)]{ralchenko08} Ralchenko Yu., Kramida A.~E., Reader J., NIST ASD Team,\ 2008, NIST Atomic Spectra Database (version 3.1.5). National Institute of Standards and Technology, Gaithersburg, MD (available online at: http://physics.nist.gov/asd, accessed on 2010 September 1)
\bibitem[Richardson et al.(2002)]{richardson02} Richardson D., Branch D., Casebeer D., Millard J., Thomas R.~C., Baron E.,\ 2002, AJ, 123, 745 
\bibitem[Roming et al.(2012)]{roming12} Roming  P.~W.~A. et al.,\ 2012, preprint (arXiv:1202.4840)
\bibitem[Schlegel, Finkbeiner \& Davis(1998)]{schlegel98} Schlegel D.~J., Finkbeiner D.~P., Davis M.,\ 1998, \apj, 500, 525 
\bibitem[Seaton(1954)]{seaton54} Seaton M.~J.,\ 1954, Ann, Astrophys., 17, 74 
\bibitem[Shields(1990)]{shields90} Shields G.~A.,\ 1990, \araa, 28, 525 
\bibitem[Sigut \& Pradhan(2003)]{sigut03} Sigut T.~A.~A., Pradhan A.~K.,\ 2003, \apjs, 145, 15 
\bibitem[Skrutskie et al.(2006)]{skrutskie06} Skrutskie M.~F. et al.,\ 2006, \aj, 131, 1163 
\bibitem[Smith et al.(2010a)]{smith10a} Smith N., Chornock R., Silverman J.~M., Filippenko A.~V., Foley R.~J.,\ 2010a, \apj, 709, 856 
\bibitem[Smith et al.(2010b)]{smith10b} Smith N. et al.,\ 2010b, \aj, 139, 1451 
\bibitem[Sollerman, Cumming \& Lundqvist(1998)]{sollerman98} Sollerman J., Cumming R.~J., Lundqvist P.,\ 1998, \apj, 493, 933 (SCL98)
\bibitem[Stanishev(2007)]{stanishev07} Stanishev V.,\ 2007, Astron. Nachr., 328, 948 
\bibitem[Stathakis \& Sadler(1991)]{stathakis91} Stathakis R.~A., Sadler E.~M.,\ 1991, \mnras, 250, 786 
\bibitem[Steele et al.(2004)]{steele04} Steele I.~A. et al.,\ 2004, \procspie, 5489, 679 
\bibitem[Steele, Cobb \& Filippenko(2009)]{steele09} Steele T.~N., Cobb B., Filippenko A.~V.,\ 2009, Cent. Bureau Electron. Telegrams, 2011, 1 
\bibitem[Thompson et al.(2009)]{thompson09} Thompson T.~A., Prieto J.~L., Stanek K.~Z., Kistler M.~D., Beacom J.~F., Kochanek C.~S.,\ 2009, \apj, 705, 1364 
\bibitem[Turatto et al.(1993)]{turatto93} Turatto M., Cappellaro E., Danziger I.~J., Benetti S., Gouiffes C., della Valle M.,\ 1993, \mnras, 262, 128 
\bibitem[Turatto, Benetti \& Cappellaro(2003)]{turatto03} Turatto M., Benetti S., Cappellaro E.,\ 2003, in Hillebrandt W., Leibundgut B., eds, Proc. ESO/MPA/MPE Workshop, From Twilight to Highlight: The Physics of Supernovae. Springer-Verlag, Berlin, p. 200 
\bibitem[Valenti et al.(2009)]{valenti09} Valenti S. et al.,\ 2009, \nat, 459, 674 
\bibitem[Valenti et al.(2011)]{valenti11} Valenti S. et al.,\ 2011, \mnras, 416, 3138 
\bibitem[Van Dyk et al.(2012)]{vandyk12} Van Dyk S.~D. et al.,\ 2012, \aj, 143, 19 
\bibitem[Vernet et al.(2011)]{vernet11} Vernet J. et al.,\ 2011, \aap, 536, A105 
\bibitem[Weaver \& Woosley(1979)]{weaver79} Weaver T.~A., Woosley S.~E.,\ 1979, \baas, 11, 724 
\bibitem[Wooden et al.(1993)]{wooden93} Wooden D.~H., Rank D.~M., Bregman J.~D., Witteborn F.~C., Tielens A.~G.~G.~M., Cohen M., Pinto P.~A., Axelrod T.~S.,\ 1993, \apjs, 88, 477 
\bibitem[Yaron \& Gal-Yam(2012)]{yaron12} Yaron O., Gal-Yam A.,\ 2012, preprint (arXiv:1204.1891) 




\end{thebibliography}
\end{document}